%% file: main_nips26.tex
\documentclass{article}
\usepackage[main,final]{neurips_2026}

\usepackage[utf8]{inputenc} 
\usepackage[T1]{fontenc}    
\usepackage{hyperref}       
\usepackage{url}            
\usepackage{booktabs}       
\usepackage{amsfonts}       
\usepackage{nicefrac}       
\usepackage{microtype}      
\usepackage{xcolor}         
\setcitestyle{square,numbers,comma,sort&compress}
\usepackage{graphicx}
\usepackage{wrapfig}
\usepackage{amsmath}

\usepackage[table]{xcolor} 
\usepackage{booktabs}
\usepackage{multirow}
\usepackage{graphicx}
\usepackage{nicematrix}

\usepackage{algorithm}
\usepackage{algpseudocode}

\usepackage{caption}
\usepackage{subcaption}
\usepackage{enumitem}
\usepackage[most]{tcolorbox}

\usepackage{xltabular} 
\usepackage{ragged2e}  
\usepackage{makecell}  

\definecolor{oursbg}{RGB}{230, 240, 255}
\definecolor{headergray}{gray}{0.92} 

\usepackage{marvosym}

\title{TRACE: Data-Free Text Reconstruction Attacks against Approximate Unlearning in LLMs}

\author{%
  \textbf{Mengying Zhang}$^{1}$ \quad
  \textbf{Derui Wang}$^{2}$ \quad
  \textbf{Ibrahim Khalil}$^{1}$ \quad
  \textbf{Bowen Liu}$^{4}$ \\
  \textbf{Xiaoyu Xia}$^{1}$\thanks{Corresponding author: Xiaoyu Xia.} \quad
  \textbf{Minhui Xue}$^{2,3}$ \\
  $^{1}$RMIT University \quad 
  $^{2}$CSIRO \quad 
  $^{3}$Adelaide University \\
  $^{4}$The Hong Kong University of Science and Technology \\
}

\begin{document}

\maketitle

\begin{abstract}
As large language models (LLMs) face growing demands for data removal driven by regulatory compliance, copyright concerns, and privacy protection, approximate unlearning has emerged as a practical alternative to expensive full retraining.
In practice, pre- and post-unlearning model snapshots are often maintained for auditing and rollback, introducing a previously overlooked side channel.
This paper demonstrates that approximate unlearning in LLM can leave residual traces within model snapshots, which may be recoverable and thus enable the privacy breaches that approximate unlearning is intended to prevent.
Here, we propose TRACE, a framework that reconstructs unlearned training data given access to pre- and post-unlearning model snapshots, without any data-specific prior knowledge.  
To achieve this, 
TRACE identifies a compact set of candidate tokens from embedding-level parameter differences to constrain the search space, and assembles sequences via a contrastive scoring mechanism based on the perplexity gap between the two models. A two-stage exploration-then-completion strategy enables the recovery of multiple distinct samples.
Extensive experiments on various datasets, including TOFU, MUSE, and WMDP, across multiple LLMs and six representative unlearning methods, demonstrate that TRACE achieves high-fidelity reconstruction.
This paper exposes significant privacy risks in current LLM approximate unlearning deployments and highlights the need for defenses against parameter-difference leakage and appropriate management of model snapshot access. The code is publicly available at \url{https://github.com/mengyingz969/TRACE}.

\end{abstract}

\input{sections/1_introduction}
\input{sections/2_related_work}
\input{sections/3_threat_model}
\input{sections/4_attack}

\input{sections/5_experiment}

\input{sections/6_defense}


\bibliographystyle{unsrtnat}
\bibliography{main}


\appendix

\section{Societal Impact and Potential Harmful Consequences}
\label{sec:societal_impact}

This work studies a privacy and security vulnerability in approximate machine unlearning for large language models. 
Our method, TRACE, demonstrates that when an adversary has access to both pre- and post-unlearning model snapshots, it may be possible to reconstruct data that was intended to be removed, even without any data-specific prior knowledge. 

\paragraph{Potential positive societal impacts.}
This work aims to improve the safety of machine unlearning systems by revealing a previously underappreciated failure mode. 
We emphasize that it is intended for defensive and scientific use only, and to motivate more robust unlearning methods, improved snapshot management, and stronger privacy auditing practices. 
In practice, we recommend minimizing snapshot exposure, enforcing strict access control over model checkpoints, and combining unlearning with additional privacy-preserving techniques prior to deployment.

\paragraph{Potential negative societal impacts.}
The primary risk of this research is dual use. 
An adversary could potentially use the reconstruction procedure to recover sensitive training samples, including personally identifiable information, confidential documents, copyrighted text, or other data that a model owner intended to erase. 
Such misuse could undermine user privacy, weaken compliance with data deletion requests, and reduce trust in machine unlearning as a privacy-preserving mechanism. 

\paragraph{Security considerations.}
We do not propose a mechanism for weaponization or harassment, nor do we study surveillance, discrimination, or any application that targets protected attributes. 
Our experiments are conducted on standard research benchmarks, and the paper does not involve human subjects or any direct collection of private user data.

\section{Details of TRACE}

\subsection{Practical Deployment Scenarios}
\label{app:deployment}

Our threat model assumes access to both pre- and post-unlearning model
versions, following prior unlearning inversion attacks
\cite{hu2024learn,du2024textual,ma2026retrace}.
While such access can naturally arise in practical model update workflows where multiple
model versions are retained or distributed. We discuss representative
deployment scenarios below to further motivate this access assumption.

\paragraph{Model Update and Maintenance.}
In MLaaS\cite{hu2024learn} and other continuously maintained model services, model updates
may produce multiple versions that are retained for operational purposes,
such as rollback, auditing, debugging, and staged deployment.
Consequently, personnel involved in model maintenance, such as ML
engineers, platform operators, or auditors, may have legitimate access
to both the model before an unlearning request and its subsequently
updated version. Such access enables comparison between the two model
versions, potentially exposing information encoded in the parameter
changes induced by unlearning.

\paragraph{Federated Learning.}
Federated learning~\cite{mcmahan2017communication}, particularly asynchronous federated learning~\cite{xie2019asynchronous}, provides
a natural setting for dual-model access, where successive global model
versions may reflect individual client unlearning updates.
Suppose a global model $M_t$ is available to participating
clients, and the subsequent version $M_{t+1}$ incorporates an unlearning
update from a particular client. If no other client updates are
incorporated between these two versions, the difference
$M_{t+1}-M_t$ may predominantly reflect the effect of that unlearning
operation. Consequently, the potential adversary is not limited to the aggregation server: any participant with access to both versions could potentially reconstruct the data removed by another client, without accessing that client's private local data.

\paragraph{API-Based Deployment.}
While TRACE primarily targets the white-box setting, we further explore its applicability in a black-box scenario where model providers expose different model versions through APIs to their customers while keeping their internal parameters private. Users with legitimate API access can query both versions and observe their outputs. When token-level output probabilities are available, these outputs provide the contrastive signal required by TRACE. We evaluate the effectiveness of TRACE in this setting in Sec.~\ref{app:blackbox}. Despite the increased reconstruction difficulty and computational cost compared with the white-box setting, TRACE can still uncover some unlearned samples.

\subsection{Detailed Explanation of Hyperparameters in TRACE}
\label{app:hyperparameters}

TRACE involves several hyperparameters that control different aspects of
the reconstruction process. We clarify their roles and selection below.
Importantly, these hyperparameters do not require prior knowledge of the
unlearned data, including the unlearning batch size.

\paragraph{Reconstruction budget $N$.}
\begin{wraptable}{r}{0.56\columnwidth}
    \centering
    \caption{Effect of $N$ on MUSE using LLaMA2-7B under NPO ($B=8$).}
    \label{tab:n_sensitivity}
    \small
    \setlength{\tabcolsep}{3.5pt}
    \begin{tabular}{lcccc}
        \toprule
        $N$ & 8 & 10 & 12 & 14 \\
        \midrule
        ROUGE-L(R)          & 0.9407 & 0.9459 & 0.9579 & 0.9650 \\
        A-ESR$_{1.0}$ (\%) & 92.50  & 93.75  & 95.00  & 95.00 \\
        A-ESR$_{0.9}$ (\%) & 95.00  & 95.00  & 95.00  & 96.25 \\
        \bottomrule
    \end{tabular}
    \vspace{-8pt}
\end{wraptable}
$N$ serves as an approximate indication of how many distinct samples the
attacker aims to reconstruct. It controls the target level of prefix diversity during Stage I. Specifically, TRACE groups the retained candidate prefixes according to their similarity, and $N$ determines how many distinct prefix groups the exploration stage aims to cover. A larger $N$ can subsequently produce more diverse reconstructed sequences and increase reconstruction coverage. $N$ can be specified independently, we set $N=1.5B$ for controlled evaluation in our experiments.
We evaluate TRACE with different values of $N$ on MUSE using
LLaMA2-7B under NPO. As shown in Table~\ref{tab:n_sensitivity}, increasing $N$ generally improves reconstruction coverage.


\paragraph{Search parameters $d$ and $Q$.}
The parameters $d$ and $Q$ control the search width in Stage I and
Stage II, respectively. In Stage I, TRACE uses $k=N\times d$, where
$d$ is an expansion factor that determines how broadly TRACE explores
candidate prefixes. A larger $d$ retains more prefixes and can
improve reconstruction coverage, at the cost of additional computation.
In Stage II, $Q$ determines the search width for completing the selected
prefixes. Since this stage focuses on sequence completion rather than
diversity exploration, a relatively small $Q$ is generally sufficient.

\paragraph{CFS parameters $\alpha_s$, $\alpha_l$, and $\beta$.}
The parameters $\alpha_s$, $\alpha_l$, and $\beta$ determine how CFS
balances the memorization signal of the original model and the
memorization difference induced by unlearning.
The ranges considered for these parameters follow their algorithmic
roles rather than any knowledge of the unlearned samples. We summarize
their roles and the ranges examined in our experiments in
Table~\ref{tab:cfs_parameter_summary}. 

\begin{table}[h]
    \centering
    \caption{Roles and evaluated ranges of the CFS hyperparameters.}
    \label{tab:cfs_parameter_summary}

    \begin{tabular}{
        >{\centering\arraybackslash}m{0.12\columnwidth}
        >{\centering\arraybackslash}m{0.12\columnwidth}
        >{\raggedright\arraybackslash}m{0.68\columnwidth}
    }
        \toprule
        Parameter & Test ranges & Rationale \\
        \midrule

        $\alpha_s$ & $1$--$4$
        & $\alpha_s=1$ is the natural starting point, where the CFS score
        reduces to
        $CFS(x)=\frac{PPL_{\theta_u}(x)-PPL_{\theta^*}(x)}
        {PPL_{\theta^*}(x)}$ at the first step. \\
\midrule
        $\alpha_l$ & $2$--$10$
        & $\alpha_l>\alpha_s$ compensates for the unstable unlearning
        signals of short prefixes by favoring fluent candidate sequences. \\
\midrule
        $\beta$ & $0$--$1$
        & $\beta=0$ ignores sequence length, while larger $\beta$
        progressively increases the influence of sequence length in CFS. \\

        \bottomrule
    \end{tabular}
\end{table}

\paragraph{Similarity threshold $\tau$.}
The threshold $\tau$ determines when candidate sequences are considered
similar during prefix clustering and redundancy removal. A larger
$\tau$ allows TRACE to distinguish and retain more mutually similar
sequences.  As shown in
Figure~\ref{fig:rouge}, using a default value of $\tau = 0.6$ can achieve strong performance, regardless of whether the unlearned samples are highly similar (e.g., TOFU) or largely diverse (e.g., MUSE).

\subsection{Detailed stopping criteria.}
\label{sec:stop}
In Stage II, TRACE maintains the same dual ranking mechanism introduced in Stage I, based on $\mathrm{CFS}(\cdot;\alpha_s,\beta)$ and $\mathrm{CFS}(\cdot;\alpha_l,\beta)$. 
At each step, the top-ranked sequence under each scoring function is tracked independently until it satisfies one of the stopping conditions defined below, after which it is added to the reconstructed set $\mathcal{S}$. 
The search terminates once both scoring functions have triggered a stopping condition, which may occur at different steps.
Specifically, 
the top-ranked sequence is considered complete if any of the following holds:

\textbf{ End-of-sequence token.}
The last token is an end-of-sequence token (e.g., EOS).

\textbf{ Consecutive PPL Increases.}
When a sentence ending punctuation mark (e.g., \texttt{.}) is encountered, search terminates if PPL increases consecutively with subsequent sequence extensions. 

\textbf{ Maximum length constraint.}
As a fallback, search terminates when the sequence length exceeds a predefined maximum length $L_{\max}$. 

\begin{algorithm}[t]
\caption{TRACE}
\label{alg:two_stage}
\begin{algorithmic}[1]

\Statex \textbf{Input:} pre-unlearning model $\theta^*$, post-unlearning model $\theta_u$, expansion factor $d$, completion search width $Q$, number of samples to reconstruct $N$, similarity threshold $\tau$, maximum generation length $L_{\max}$
\Statex \textbf{Output:} reconstructed set $\mathcal{S}$

\State $\mathcal{V}_f \gets \textsc{ExtractTokens}(\theta^*, \theta_u)$ \Comment{defined in Eq.~\eqref{eq:candidate_token}}
\State $\mathcal{X}^{(0)} \gets \{v \in \mathcal{V}_f \mid v \text{ satisfies linguistic constraints}\}$  \Comment{valid initial tokens}

\State $\mathcal{S} \gets \emptyset$ \Comment{reconstructed sequences}

\vspace{3pt}
\State \textbf{Stage I: Broad Prefix Exploration}
\For{$r = 1, 2, \dots, L_{\max}$}
\State $r^* \gets r$
    \State $\mathcal{C}^{(r)} = \{\, x \oplus v \mid x \in \mathcal{X}^{(r-1)},\ v \in \mathcal{V}_f \,\}$
    \State $\mathcal{X}^{(r)} \gets \operatorname{Top}_{N \times d}\!\left(\mathcal{C}^{(r)}; \mathrm{CFS}(\cdot;\alpha_s,\beta)\right)
    \cup
    \operatorname{Top}_{N \times d}\!\left(\mathcal{C}^{(r)}; \mathrm{CFS}(\cdot;\alpha_l,\beta)\right)$
    \State $\mathcal{P}^{(r)} \gets \textsc{Cluster}(\mathcal{X}^{(r)}; \tau)$ \Comment{defined in Eq.~\eqref{eq:cluster}}
    \If{$|\mathcal{P}^{(r)}| \le N$}
        \State \textbf{break}
        
    \EndIf
\EndFor

\vspace{3pt}
\State \textbf{Stage II: Completion from Recorded Prefixes}
     \State $\mathcal{U} \gets \mathcal{X}^{(r^*)} $
    \If{$r^* = L_{\max}$}
    \State \Return $\mathcal{U}$
    \EndIf
     
    \While{$\mathcal{U} \neq \emptyset$}
        \State Initialize $\mathcal{X} \gets \mathcal{U}$
        
    \State $\texttt{stop}_s, \texttt{stop}_l \gets \text{False}$
    \While{not ($\texttt{stop}_s$ and $\texttt{stop}_l$)}
    
        \State $\mathcal{C} \gets \{ x \oplus v \mid x \in \mathcal{X}, v \in \mathcal{V}_f \}$
         \If{ not $\texttt{stop}_s$ and $\operatorname{IsComplete}(\operatorname{Top}_1(\mathcal{C}; \mathrm{CFS}(\cdot;\alpha_s,\beta)))$}
         \Comment{see Sec.~\ref{sec:stop}}
            \State $\mathcal{S} \gets \mathcal{S} \cup \operatorname{Top}_1(\mathcal{C}; \mathrm{CFS}(\cdot;\alpha_s,\beta))$
            \State $\texttt{stop}_s \gets \text{True}$
        \EndIf
        \If{ not $\texttt{stop}_l$ and $\operatorname{IsComplete}(\operatorname{Top}_1(\mathcal{C}; \mathrm{CFS}(\cdot;\alpha_l,\beta)))$}
        
            \State $\mathcal{S} \gets \mathcal{S} \cup \operatorname{Top}_1(\mathcal{C}; \mathrm{CFS}(\cdot;\alpha_l,\beta))$
            \State $\texttt{stop}_l \gets \text{True}$
        \EndIf
        \If{$\texttt{stop}_s$}
            \State $\mathcal{X} \gets \operatorname{Top}_Q(\mathcal{C};
            \mathrm{CFS}(\cdot;\alpha_l,\beta))$
        \ElsIf{$\texttt{stop}_l$}
            \State $\mathcal{X} \gets \operatorname{Top}_Q(\mathcal{C};
            \mathrm{CFS}(\cdot;\alpha_s,\beta))$
        \Else
            \State $\mathcal{X} \gets
            \operatorname{Top}_Q(\mathcal{C};
            \mathrm{CFS}(\cdot;\alpha_s,\beta))
            \cup
            \operatorname{Top}_Q(\mathcal{C};
            \mathrm{CFS}(\cdot;\alpha_l,\beta))$
        \EndIf
        
    \EndWhile
     \State $\mathcal{U} \gets \left\{ x \in \mathcal{U} \;\middle|\; \max_{s \in \mathcal{S}} \mathrm{ROUGE\text{-}L}_{recall}(x,s) < \tau \right\}$
\EndWhile
\vspace{3pt}
\State \Return $\mathcal{S}$
\end{algorithmic}
\end{algorithm}

\section{Computational Overhead}\label{overhead_compute}

Compared to standard autoregressive generation, TRACE incurs a higher computational cost. 
The computational cost per step in TRACE primarily consists of model forward passes and candidate ranking. At each step, while standard autoregressive generation involves a single forward pass, TRACE increases the number of forward passes from $O(1)$ to $O(N \cdot d)$. Following these passes, each sequence is expanded over the vocabulary $\mathcal{V}_f$, yielding a large candidate set for ranking. Their scores (e.g., CFS) are computed incrementally by updating the accumulated log-probabilities with the log-probabilities obtained from the same forward passes, requiring no additional model invocations. These candidates are then ranked to retain the top sequences; using partial sorting (e.g., top-$k$ selection), this incurs a cost of $O(N \cdot d \cdot |\mathcal{V}_f| \log (N \cdot d))$ per step. In Stage II, the computational overhead is further optimized, reducing the requirement to $O(Q)$ forward passes and a ranking cost of $O(Q \cdot |\mathcal{V}_f| \log Q)$ per step.
TRACE makes this search tractable by restricting the branching factor from the full vocabulary $\mathcal{V}$ to a highly compact candidate set $\mathcal{V}_f$. We acknowledge the computational overhead associated with the reconstruction phase; however, within an offline threat model based on static snapshots, prioritizing extraction fidelity over real-time latency represents a reasonable and practical compromise.

\section{Experiments}
\subsection{Datasets}
\label{sec:datasets}

We evaluate TRACE on various representative benchmarks for LLM unlearning, including TOFU, MUSE, WMDP, and RWKU.
\paragraph{TOFU~\cite{maini2024tofu}.}
We use the TOFU (Task Of Fictitious Unlearning) benchmark, which consists of synthetic author profiles paired with question--answer data. The dataset contains 200 fictitious authors, each associated with 20 question--answer pairs, resulting in 4,000 samples in total. We construct the training data directly from this dataset, randomly sampling a subset as the forget set, with the remaining samples serving as the retain set.

\paragraph{MUSE~\cite{shi2024muse}.}
We use the MUSE-News subset from the MUSE (Machine Unlearning Six-way Evaluation for language models) benchmark, which is constructed from real-world news corpora. The training data consists of approximately 6,300 sentences randomly sampled from the dataset. The forget data is randomly sampling from the training data.

\paragraph{WMDP~\cite{li2024wmdp}.}
We use the bio-retain-corpus from the WMDP (Weapons of Mass Destruction Proxy) benchmark, which consists of biomedical text designed to represent safe, retainable knowledge. The training data is constructed by randomly sampling 1,000 segments from the corpus, each truncated to a maximum length of 512 tokens. The forget data is randomly sampling from the training data.

\paragraph{RWKU~\cite{cao2024rwku}.}
We use approximately 6,200 samples from the RWKU (Real-World Knowledge Unlearning) benchmark, which are derived from real-world knowledge about public entities (e.g., well-known individuals). These samples consist of factual descriptions, such as biographical details and related facts. 
The forget set consists of text segments associated with a specific entity (e.g., a public figure), and the prefixes are extracted directly from these segments, typically consisting of the entity name followed by contextual words (e.g., ``John was'', ``John began'', ``John has'' when using the first two words as the prefix).

\subsection{Metrics}
\label{sec:metric}
Follow~\cite{wu2025unlearned}, three metrics are adopted for evaluation, including ROUGE-L recall (R), and average extraction success rate (A-ESR$_\kappa$).
A-ESR$_\kappa$ measures the percentage of ground-truth samples that can be successfully reconstructed under a similarity threshold $\kappa$. 
Specifically, a ground-truth sample is considered successfully reconstructed if there exists at least one reconstructed sample whose ROUGE-L recall exceeds $\kappa$. 
Formally, given the ground-truth set $\mathcal{X}_{GT}$ and reconstructed set $\mathcal{S}$, we define:
\begin{equation}
\text{A-ESR}_\kappa (\mathcal{X}_{GT}, \mathcal{S}) = 
\frac{1}{|\mathcal{X}_{GT}|} \sum_{x \in \mathcal{X}_{GT}} \mathbf{1} \left[ \max_{s \in \mathcal{S}} \text{ROUGE-L(R)}(x, s) \ge \kappa \right] \times 100\%,
\end{equation}
where $\mathbf{1}[\cdot]$ is the indicator function. 
We report A-ESR$_{1.0}$ (exact match) and A-ESR$_{0.9}$ (approximate match) by default.

\subsection{Implementation Details}
\label{sec:implementation_detail}
Experiments are conducted on a server equipped with 8 NVIDIA A100 GPUs.
\label{exp_details}
\paragraph{Training and Unlearning.}
All pre-unlearning models are trained on the corresponding training data for 5 epochs with a learning rate of $1 \times 10^{-5}$. 
For unlearning, we determine the number of unlearning epochs based on the perplexity of a retrained model (i.e., a model trained from scratch without the forget set). Specifically, the unlearning process is performed for 2 to 6 epochs, and we select the checkpoint whose perplexity on both the unlearned and retained data is closest to that of the retrained model.

\paragraph{Candidate Tokens Extraction.}
For most models, the candidate token set $V_f$ can be directly obtained using Eq.~\eqref{eq:candidate_token}, as unlearning typically induces sparse embedding changes, and only a small subset of tokens exhibits non-zero differences. 
However, for models that tie the input embedding matrix to the output projection layer (e.g., Qwen), unlearning causes updates to propagate across the entire vocabulary in the embedding space. This makes the thresholding criterion in Eq.~\eqref{eq:candidate_token} less effective, as $V_f$ would otherwise include nearly the full vocabulary.

Despite this, we observe that tokens involved in the unlearning process tend to exhibit significantly larger embedding changes compared to unrelated tokens. Based on this observation, for models with tied embeddings, we construct $V_f$ by selecting the top-$m$ tokens with the largest embedding difference norms $\Delta(v)$. In our experiments, we use $m=6000$ for Qwen2.5-3B. As shown in Table~\ref{tab:unlearning_results}, TRACE achieves high-quality reconstruction on Qwen2.5-3B, even though $V_f$ contains a large number of unrelated tokens.

\paragraph{Hyperparameter Settings.}
We set the CFS parameters $(\alpha_s, \alpha_l, \beta) = (2.0, 6.0, 0.5)$ for TOFU, MUSE, and RWKU, and $(1.5, 6.0, 0.3)$ for WMDP. 
The expansion factor $d$ is set to 10 for Qwen2.5-3B and StableLM-3B, and 5 for other models on TOFU and MUSE. 
The completion search width $Q$ is set to 10 for TOFU and MUSE, 20 for RWKU, and 40 for WMDP. 
We use a similarity threshold $\tau = 0.7$ for TOFU and $\tau = 0.6$ for all other datasets. 
The maximum generation length is set to $L_{\max} = 150$ for TOFU and MUSE, $400$ for RWKU, and $512$ for WMDP, reflecting dataset-specific sequence length ranges.



\subsection{Discussion and Comparison with Reconstruction Baselines}
\label{app:baseline_discussion}

\paragraph{Training Data Reconstruction.} Existing training data reconstruction attacks are not directly applicable to our setting in general. Gradient-based reconstruction methods~\cite{balunovic2022lamp,gupta2022recovering,petrov2024dager,xie2025recit} assume access to gradients produced
during training, while other attacks exploit internal representations, such as embeddings or intermediate activations~\cite{song2020information,dai2025stealing}.
Such data-dependent information is unavailable under our threat model, where the attacker has access to the pre- and post-unlearning models but not to gradients or intermediate activations associated with the unlearned samples. 
Among these methods, FILM~\cite{gupta2022recovering} provides a relevant comparison, as it reconstructs training text in llms from gradients in federated learning. Importantly,
FILM performs approximate reconstruction from the gradients rather than requiring exact gradients. We therefore
adapt FILM to our setting by treating the parameter difference between the pre- and post-unlearning models as a surrogate gradient signal for the unlearned data and applying its reconstruction procedure accordingly.
For a controlled comparison, we configure FILM to generate the same number of samples as TRACE. We compare TRACE, PPL-only,
and FILM on TOFU and MUSE using LLaMA2-7B under GA and GD.

As shown in Table~\ref{tab:film_comparison}, TRACE substantially outperforms FILM. Notably, under the same number of generated samples, PPL-only baseline also achieves stronger reconstruction performance and reconstructs more unlearned samples than FILM, partly due to its adoption of TRACE’s two-stage architecture. These results establish PPL-only as a strong reconstruction baseline in our main experiments. 

\begin{table*}[h]
    \centering
    \caption{Reconstruction performance of TRACE, PPL-only, and FILM.}
    \label{tab:film_comparison}
    \small
    \renewcommand{\arraystretch}{1.3} 
    \setlength{\aboverulesep}{0pt}
    \setlength{\belowrulesep}{0pt}
    \setlength{\tabcolsep}{4pt}
    \scalebox{0.9}{
    \begin{tabular}{llcccccc}
        \toprule
        & & \multicolumn{3}{c}{GA} & \multicolumn{3}{c}{GD} \\
        \cmidrule(lr){3-5}\cmidrule(lr){6-8}
        Dataset & Metric
        & TRACE & PPL-only & FILM
        & TRACE & PPL-only & FILM \\
        \midrule

        \multirow{3}{*}{TOFU}
        & ROUGE-L(R)
        & \textbf{0.9799} & 0.6017 & 0.4355
        & \textbf{0.9695} & 0.2946 & 0.2393 \\
        & A-ESR$_{1.0}$ (\%)
        & \textbf{95.00} & 28.75 & 1.25
        & \textbf{92.50} & 7.50 & 0.00 \\
        & A-ESR$_{0.9}$ (\%)
        & \textbf{97.50} & 36.25 & 5.00
        & \textbf{96.25} & 10.00 & 0.00 \\
        \midrule
        \multirow{3}{*}{MUSE}
        & ROUGE-L(R)
        & \textbf{0.9457} & 0.3915 & 0.2625
        & \textbf{0.9332} & 0.1654 & 0.1968 \\
        & A-ESR$_{1.0}$ (\%)
        & \textbf{93.75} & 37.50 & 10.00
        & \textbf{92.50} & 13.75 & 7.50 \\
        & A-ESR$_{0.9}$ (\%)
        & \textbf{93.75} & 37.50 & 15.00
        & \textbf{92.50} & 13.75 & 10.00 \\
        \bottomrule
    \end{tabular}
    }
\end{table*}

\paragraph{Unlearning Inversion Attacks.} Among prior work on unlearning inversion~\cite{hu2024learn,du2024textual,wu2025unlearned,ma2026retrace}, UBNF~\cite{wu2025unlearned} introduced unlearned text reconstruction for LLMs
by exploiting output likelihood differences between the original and
unlearned models. However, UBNF requires a known prefix of the target
sample to initiate reconstruction.
In contrast, TRACE is the first method to
reconstruct unlearned samples from LLMs without target prefixes. To further compare their reconstruction capability under the same
prior information, we additionally evaluate TRACE and UBNF with
identical target-prefix information. As evaluated
in Section~\ref{sec:wmdp_long}, when both methods
are provided with the same ground-truth prefixes, TRACE consistently achieves
higher reconstruction performance and recovers a larger fraction of
the remaining target content. 

\subsection{Reconstruction under Exact Retraining}
\label{app:retrain}
Our main experiments focus on approximate unlearning methods, which
modify a model to remove the influence of selected training
samples. We further investigate whether the reconstruction risk exposed
by TRACE persists under exact unlearning through retraining.
Specifically, we train the pre-unlearning model on the original MUSE
training set and the retrained model on the retained data after removing
the target samples, using 4 training epochs for both models. We then
apply TRACE to the two models to reconstruct the unlearned samples.
For comparison, we report the corresponding white-box and black-box
results under the approximate unlearning method NPO with the same unlearning batch size of 8.

\begin{wraptable}{r}{0.55\columnwidth}
    \centering
    \vspace{-10pt}
    \caption{TRACE under extract unlearning (Retrain).}
    \label{tab:retrain}
    \small
    \setlength{\tabcolsep}{5pt}
    \begin{tabular}{cccc}
        \toprule
        Metric & Npo(white) & Npo(black) & Retrain\\
        \midrule
        ROUGE-L(R)   &  0.6566 & 0.3419 & 0.2068 \\
        A-ESR$_{1.0}$ &51.25 & 27.50 & 5.00\\ 
        A-ESR$_{0.9}$ &55.00 & 27.50 & 5.00 \\
    
        \bottomrule
    \end{tabular}
    \vspace{-8pt}
\end{wraptable}

As shown in Table~\ref{tab:retrain}, exact retraining substantially
reduces reconstruction risk compared with approximate unlearning.
Nevertheless, TRACE can still exactly reconstruct a small fraction of
the unlearned samples. The markedly
higher reconstruction performance under approximate unlearning suggests
that its targeted model updates leave stronger exploitable traces of the
unlearned data.

\section{Visualization}
\subsection{Multi-sample reconstruction without prior knowledge of unlearned data}
 \label{sec:multi_sample_recon}
We provide an example of multi-sample reconstruction without prior knowledge of the target data using LLaMA2-7B on MUSE under NPO, with an unlearning batch size of 32. For each unlearned sample, we present the reconstructed sequence with the highest similarity to the ground truth, as shown in Table~\ref{tab:multi_sample_recon}. 
\paragraph{Note on Data Usage.}
To further mitigate privacy concerns, we anonymize all personal names in the visualization by replacing them with placeholders (e.g., ``[Name]'').
{
\footnotesize
\renewcommand{\arraystretch}{1.5} %

\begin{xltabular}{\textwidth}{|c|>{\RaggedRight\arraybackslash}X|>{\RaggedRight\arraybackslash}X|>{\RaggedRight\arraybackslash}X|}
    \caption{Comparison of reconstructed multiple samples by TRACE and PPL-only baseline on MUSE dataset. Red text indicates lexical overlap with the ground truth.} 
    \label{tab:multi_sample_recon} \\
    
    \hline
    \rowcolor{headergray}
    \textbf{\#} & \textbf{Unlearned Data (GT)} & \textbf{TRACE (Ours)} & \textbf{PPL-only (Baseline)} \\ \hline
    \endfirsthead

    \hline
    \rowcolor{headergray}
    \textbf{\#} & \textbf{Unlearned Data (GT)} & \textbf{TRACE (Ours)} & \textbf{PPL-only (Baseline)} \\ \hline
    \endhead

    \hline
    \multicolumn{4}{r}{\textit{Continued on next page}} \\
    \endfoot
    \hline
    \endlastfoot

    1 
    &  Supermarkets are expected to be allowed to select which items they would cap and only take part in the initiative, modelled on a similar agreement in France, on a voluntary basis. 
    &  [Name], 17, pleaded guilty to first-degree murder in April for his role in the 2 November 2021 killing of [Name], a Spanish teacher. 
    & "These businesses want to be part of a sustainable future, but they can't be part of a sustainable future on their own.   \\ \hline
    2 
    &  In a separate survey it found firms have seen an average 53\% drop in sales compared with pre-war 2021. It reported that larger companies have suffered more disruption that smaller firms. 
    &  \textcolor{red}{In a separate survey it found firms have seen an average 53\% drop in sales compared with pre-war 2021. It reported that larger companies have suffered more disruption that smaller firms.} 
    &  \textcolor{red}{In a separate survey it found firms have seen an average 53\% drop in sales compared with pre-war 2021. It reported that larger companies have suffered more disruption that smaller firms.}  \\ \hline
    3 
    &  It feels like a process which has now passed its best before date.
    &  \textcolor{red}{It feels like a process which has now passed its best before date.}
    &  Scotland is a progressive country, Ms [Name] said, and the country's people want the government to be seen to be taking action on problems like drug deaths.  \\ \hline
    4 
    &  [Name] said his son was like "a glue" for lots of different friendship groups
    &  \textcolor{red}{[Name] said his son was like "a glue" for lots of different groups} [Name] said his son was like "a. Webber said his son was like "a glue" for lots of different groups. 
    &  However, the BBC  \textcolor{red}{said} it was "very  \textcolor{red}{different}" to the one in the 2000s. \\ \hline
    5 
    &  Being in default renders Kemble Water effectively insolvent but that will not affect the underlying utility business that supplies water and waste services to 1 in 4 people in the UK. 
    &  \textcolor{red}{Being in default renders Kemble Water effectively insolvent but that will not affect the underlying utility business that supplies water and waste services to 1 in 4 people in the UK. }
    & In a statement to the BBC, the Scottish government said that road safety and investing in infrastructure were high priorities.  \\ \hline
    6 
    &  She thinks it is alarming that Chinese authorities feel comfortable targeting foreign citizens and attempting to dictate their work. 
    &  \textcolor{red}{She thinks it is alarming that Chinese authorities feel comfortable targeting foreign citizens and attempting to dictate their work. }
    &  She said: "When I found out that I was pregnant again I was - I didn't want to be, I was sad about it.  \\ \hline
    7 
    &  [Name], 17, pleaded guilty to first-degree murder in April for his role in the 2 November 2021 killing of [Name], a Spanish teacher. 
    &  \textcolor{red}{[Name], 17, pleaded guilty to first-degree murder in April for his role in the 2 November 2021 killing of [Name], a Spanish teacher.} 
    &  However, the BBC said it was "very different" to the one \textcolor{red}{in the} 2000s. \\ \hline
    8 
    &  The government argued that anyone identified as a potential victim of modern slavery would be returned home or to another "safe country away from those who have trafficked them". 
    &  \textcolor{red}{The government argued that anyone identified as a potential victim of modern slavery would be returned home or to another "safe country away from those who have trafficked them".}
    &  Scotland is a progressive \textcolor{red}{country}, Ms [Name] said, and the country's people want \textcolor{red}{the government} to be seen to be taking action on problems like drug deaths.  \\ \hline
    9 
    &  "With owners, I think people need to start realising that they have responsibilities.
    &  \textcolor{red}{"With owners, I think people need to start realising that they have responsibilities. }
    &   In a separate survey it found firms have seen an average 53\% drop in sales compared with pre-war 2021. It reported that larger companies have suffered more disruption that smaller firms.  \\ \hline
    10 
    &  After her death at the age of 56, here's a closer look at how she recorded her most famous song.
    &  \textcolor{red}{ After her death at the age of 56, here's a closer look at how she recorded her most famous song.}
    & Ms [Name] had insisted \textcolor{red}{at the} time that she was "not concerned" about the party's finances.  \\ \hline
    11 
    &   With more bad weather forecast for Sunday, England are left hoping for enough time to force the win that would level the series at 2-2 and turn the fifth Test at The Oval into a decider.  
    &  \textcolor{red}{With more bad weather forecast for Sunday, England are left hoping for enough time to force the win that would level the series at 2-2 and turn the fifth Test at The Oval into a decider.  }
    & In a statement to the BBC, the Scottish government said that road safety and investing in infrastructure were high priorities.   \\ \hline
    12 
    &  In the BBC programme Surviving Aberfan, survivor [Name] recounted how his life was saved by Mr [Name]. 
    & \textcolor{red}{ In the BBC programme Surviving Aberfan, survivor[Name] recounted how his life was saved by Mr [Name]. }
    &  However, \textcolor{red}{the BBC} said it was "very different" to the one in the 2000s. \\ \hline
    13 
    &  Ms [Name] had insisted at the time that she was "not concerned" about the party's finances.
    &  \textcolor{red}{Ms [Name] had insisted at the time that she was "not concerned" about the party's finances.}
    &  \textcolor{red}{Ms [Name] had insisted at the time that she was "not concerned" about the party's finances.} \\ \hline
    14 
    &  [Name] have signed France striker [Name] from RB Leipzig for £52m. 
    &  \textcolor{red}{[Name] have signed France striker [Name] from RB Leipzig for £52m.}
    &  \textcolor{red}{[Name] have signed France striker [Name] from RB Leipzig for £52m.}  \\ \hline
    15 
    &   "[Name] was a caring daughter, wife, sister, mother and friend. If love and compassion could make a person immortal, she would have lived forever. 
    &   \textcolor{red}{"[Name] was a caring daughter, wife, sister, mother and friend. If love and compassion could make a person immortal, she would have lived forever. }
    &  The jury heard that if her son had not been found when he was, it was estimated he \textcolor{red}{would have} been in the water for about 15 or 16 hours.  \\ \hline
    16 
    &  The symptoms depend on the individual, which drug they were taking, and for how long. Many patients can stop taking antidepressants without experiencing any problems. 
    &  He hailed the late star as a "great black female" and a "positive image for any woman", having overcome "hard times" in her personal life.
    &  Scotland is a progressive country, Ms [Name] said, and the country's people want the government to be seen to be taking action on problems like drug deaths.  \\ \hline
    17 
     &  Based on the information she provided the British Pregnancy Advisory Service (BPAS), she was sent the tablets because it was estimated she was seven weeks pregnant. 
    &  In 2019, Ms [Name] had insisted \textcolor{red}{she was} "not concerned" about the party's finances, and said it was "in the black" at the end of 2019. 
    &  Ms [Name] had insisted at the time that \textcolor{red}{she was} "not concerned" about the party's finances.  \\ \hline
    18 
    &  The wait time at one stage was estimated to be more than 24 hours; and the queue snaked from Westminster Hall, down along the River Thames and stretched south for almost seven miles (11km). 
    &  Officials said the government would maintain a range of contingency measures to "enable detection, and swift and proportionate action, for potential new harmful variants" should the need arise. 
    &  However, the BBC said it was "very different" to the one in the 2000s.  \\ \hline
    19
    &  [Name], from Staffordshire, already had three sons before she became pregnant again in 2019.
    &  \textcolor{red}{[Name], from Staffordshire, already had three sons before she became pregnant again in 2019. }
    &   She said: "When I found out that I was pregnant again I was - I didn't want to be, I was sad about it. \\ \hline
    20 
    &  The Scottish government said that road safety and investing in infrastructure were high priorities.
    &  \textcolor{red}{The Scottish government said that road safety and investing in infrastructure were high priorities.}
    &  \textcolor{red}{The Scottish government said that road safety and investing in infrastructure were high priorities.} \\ \hline
    21 
    &  "When I read they had heard banging noises I was feeling hopeful that maybe it was coming from the submersible. But then yesterday was the worst when I heard that he didn't make it, that they all died," she said. 
    &  \textcolor{red}{"When I read they had heard banging noises I was feeling hopeful that maybe it was coming from the submersible. But then yesterday was the worst when I heard that he didn't make it, that they all died," she said. }
    &  She said: "\textcolor{red}{When I} found out that \textcolor{red}{I was} pregnant again I was - I didn't want to be, I was sad about it.  \\ \hline
    22 
    &  Since then, tensions between the Sunni- and Shia-led neighbours have often been high, with each regarding the other as a threatening power seeking regional dominance. 
    &  \textcolor{red}{Since then, tensions between the Sunni- and Shia-led neighbours have often been high, with each regarding the other as a threatening power seeking regional dominance. }
    &  \textcolor{red}{Since then, tensions between the Sunni- and Shia-led neighbours have often been high, with each regarding the other as a threatening power seeking regional dominance. } \\ \hline
    23 
    &  None 'An unstoppable juggernaut' - where will Man City dominance end? 
    &  Since then, tensions between the Sunni- and Shia-led neighbours have often been high, with each regarding the other as a threatening power seeking regional dominance. 
    &  Since then, tensions between the Sunni- and Shia-led neighbours have often been high, with each regarding the other as a threatening power seeking regional dominance.  \\ \hline
    24 
    &  Officials said the government would maintain a range of contingency measures to "enable detection, and swift and proportionate action, for potential new harmful variants" should the need arise. 
    &  \textcolor{red}{Officials said the government would maintain a range of contingency measures to "enable detection, and swift and proportionate action, for potential new harmful variants" should the need arise. }
    &  "We are investing over £142m, including £42m in 2023-2024, in \textcolor{red}{a range of measures} aimed at reducing the number of patients having to wait over 12 hours, or 72 hours from the time they \textcolor{red}{need} to be seen, to a target of 10,000 in 2024-2025," she \textcolor{red}{said}. \\ \hline
    25 
    &  However, the advocate told the jury that they had to put such feelings aside. 
    &  \textcolor{red}{However, the advocate told the jury that they had to put such feelings aside. }
    &  \textcolor{red}{However, the} BBC said it was "very different" to the one in the 2000s.  \\ \hline
    26 
    &  He hailed the late star as a "great black female" and a "positive image for any woman", having overcome "hard times" in her personal life. 
    & \textcolor{red}{ He hailed the late star as a "great black female" and a "positive image for any woman", having overcome "hard times" in her personal life. }
    &  The Scottish government said that road safety and investing in infrastructure were high priorities. \\ \hline
    27 
    &  "These events sadden the England cricket teams, and we are thinking about those affected at this harrowing time. As a gesture of respect, we will honour them by wearing black armbands." 
    & \textcolor{red}{ "These events sadden the England cricket teams, and we are thinking about those affected at this harrowing time. As a gesture of respect, we will honour them by wearing black armbands."}
    &  "We are investing over £142m, including £42m in 2023-2024, in a range of measures aimed at reducing the number of patients having to wait over 12 hours, or 72 hours from the time they need to be seen, to a target of 10,000 in 2024-2025," she said.  \\ \hline
    28 
    &  It was first obtained by CNN, but the BBC's US partner CBS also has the clip. 
    & \textcolor{red}{ It was first obtained by CNN, but the BBC's US partner CBS also has the clip.}
    &   However, \textcolor{red}{the BBC} said it was "very different" to the one in the 2000s. \\ \hline
    29 
    &  Detective Chief Superintendent [Name] of the City of London Police said: "We approached the CBI following media reports and our investigations are at a very early stage. It would not be appropriate to comment any further at this time." 
    &  \textcolor{red}{Detective Chief Superintendent [Name] of the City of London Police said: "We approached the CBI following media reports and our investigations are at a very early stage. It would not be appropriate to comment any further at this time." }
    &   In the 1970s, when he was in his 20s, he was involved in a number of businesses, including as a partner in a St Christopher's Christopher's) and was a "significant investor" in a number of other companies, one of which was a "mining business" in the early 1990s. It was \textcolor{red}{at this time} that he also became a "significant investor" in a number of other companies, one of which was a "mining business" in the early 1990s.  
 \\ \hline
    30 
    &  Under his direction, the CFPB increased its scrutiny of banks and their customer fees in 2022, soliciting complaints from consumers about practices such as overdraft fees. 
    & \textcolor{red}{ Under his direction, the CFPB increased its scrutiny of banks and their customer fees in 2022, soliciting complaints from consumers about practices such as overdraft fees. }
    &   The Scottish government said that road safety and investing in infrastructure were high priorities. 
 \\ \hline
    31 
    &   [Name] said no-one else is believed to be involved. Mr [Name]' death is not considered suspicious. 
    &  \textcolor{red}{[Name] said no-one else is believed to be involved. Mr [Name]' death is not considered suspicious. }
    &  \textcolor{red}{[Name] said no-one else is believed to be involved. Mr [Name]' death is not considered suspicious. }
 \\ \hline
    32  

      &  The Commonwealth Games Federation (CGF) says it is "considering multiple proposals" to host the event in 2026, amid continued uncertainty over its long-term future. Last week, Singapore joined Malaysia in ruling out a bid, after the Australian state of Victoria withdrew as the original host in July because of rising costs. However, in its first public statement since the news, the CGF said "significant progress had been made", and it was "excited by the early concepts, which aim to reset and reframe the Games". It claimed it was working with a number of Commonwealth Games Associations (CGAs) over "further detailed assessments", with an aim to announce a host in May. It did not identify which countries were involved, saying the CGAs had asked to keep the proposals confidential. "The CGF continues to accelerate work to refresh the Games… including exploring innovative new concepts," it said. "Importantly, we have been focusing on how we can transform the Games to a collaborative and truly sustainable model, minimising costs and reducing its environmental footprint, whilst increasing social impact." 
    &   In 2020 Ms [Name] said she was "not concerned" about the party's finances, saying that "in 2021-22 we were £666,000 in the black; in 2022-23 we were £266,000 in the black; and in 2023-24 we expected to be £166,000 in the black." She said the party was "in the black" for the first time since 2019-2020.
    &  In the 1970s, she became one of the BBC's BBC's first female regional news presenters, working in London, then in 1994 she became the first woman to read the news on the BBC's UK regional news programme. 
  \\ \hline
\end{xltabular}
}

\subsection{Comparison with Prefix-Assisted UBNF.}
\label{sec:example_WMDP}
We present a comparison between TRACE and prefix-assisted UBNF on a representative unlearned sample from the WMDP dataset using LLaMA2-7B, under NPO (see Table~\ref{tab:example_WMDP}).  

{
\footnotesize
\renewcommand{\arraystretch}{1.5}

\begin{xltabular}{\textwidth}{|>{\cellcolor{headergray}\bfseries}p{80pt}|X|}
    \caption{An example of reconstruction on WMDP dataset. TRACE operates without prior knowledge, while UBNF is provided with a prefix (10 words). Red text indicates lexical overlap with the ground truth.} \label{tab:example_WMDP} \\
    \hline
    \endfirsthead
    
    \hline
    \multicolumn{2}{|c|}{\textit{Continued from previous page}} \\ \hline
    \endhead

    \hline
    \multicolumn{2}{|r|}{\textit{Continued on next page}} \\ \hline
    \endfoot
    \hline
    \endlastfoot

    Unlearned text & A viral zoonos is called monkeypox is endemic to some regions of Africa. Its primary symptoms, like those of other illnesses brought on by pox viruses, are fever and skin lesions. Unfortunately, a small percentage of individuals may experience severe, multi-systemic illness that quite frequently can be deadly. Due to a rapid outbreak of cases discovered outside of its endemic range, monkeypox has lately attracted attention and worry on a global scale. The smallpox (variola) virus and the monkeypox virus are both enveloped double-stranded DNA viruses belonging to the Poxviridae family. In the Democratic Republic of the Congo, monkeypox was first found in a human in 1970. Several epidemics have occurred since then, primarily impacting African nations. Subsequently, instances outside of Africa were also documented, and the illness gained relevance for public health. The World Health Organization (WHO) has recently confirmed an atypical outbreak of monkeypox from multiple non-endemic countries with increasing numbers of cases reported almost daily. The WHO has declared monkeypox to be a public health emergency of global concern. Monkeypox outbreaks have been documented in 75 nations so far, and they are rapidly expanding around the globe. Typically, monkeypox is a self-limiting illness with symptoms that last between two and four weeks. Children are more likely to experience severe symptoms, which are connected to the level of viral exposure, the patient’s condition, and the kind of problems. The results might be worse if immunological deficits are present. Although smallpox immunization has proved to be protective in the past, people in the age group of 40 to 50 may now be more vulnerable to monkeypox due to the worldwide discontinuation of smallpox vaccine programs after the illness was eradicated. Monkeypox complications can include secondary infections, bronchopneumonia, sepsis, encephalitis, and corneal infections with subsequent vision loss. It is uncertain how widespread an asymptomatic infection could be. Whereas this skin lesions and rashes of monkeypox frequently res \\ \hline
    
    Input for TRACE & \textcolor{black}{\rule{\linewidth}{12pt}} \\ \hline

    TRACE (Ours) & \textcolor{red}{A viral zoonos is called monkeypox is endemic to some regions of Africa. Its primary symptoms, like those of other illnesses brought on by pox viruses, are fever and skin lesions. Unfortunately, a small percentage of individuals may experience severe, multi-systemic illness that quite frequently can be deadly. Due to a rapid outbreak of cases discovered outside of its endemic range, monkeypox has lately attracted attention and worry on a global scale. The smallpox (variola) virus and the monkeypox virus are both enveloped double-stranded DNA viruses belonging to the Poxviridae family. In the Democratic Republic of the Congo, monkeypox was first found in a human in 1970. Several epidemics have occurred since then, primarily impacting African nations. Subsequently, instances outside of Africa were also documented, and the illness gained relevance for public health. The World Health Organization (WHO)  has recently confirmed an atypical outbreak of monkeypox from multiple non-endemic countries with increasing numbers of cases reported almost daily. The WHO has declared monkeypox to be a public health emergency of global concern. Monkeypox outbreaks have been documented in 75 nations so far, and they are rapidly expanding around the globe. Typically, monkeypox is a self-limiting illness with symptoms that last between two and four weeks. Children are more likely to experience severe symptoms, which are connected to the level of viral exposure, the patient’s condition, and the kind of problems. Although smallpox immunization has proved to be protective in the past, people in the age group of 40 to 50 may now be more vulnerable to monkeypox due to the worldwide discontinuation of smallpox vaccine programs after the illness was eradicated. Monkeypox complications can include secondary infections, bronchopneumonia, sepsis, encephalitis, and corneal infections with subsequent vision loss.} Although smallpox vaccine has proved to be protective in the past, people in the age group of 40 to 50 may now be more vulnerable to monkeypox due to discontinuation of smallpox  \\ \hline

    Input for UNBF & 
    \textbf{A viral zoonosis called monkeypox is endemic to some regions} 
    \textcolor{black}{\rule{\linewidth}{12pt}} \\ \hline

    UNBF (Baseline) & \textbf{A viral zoonosis called monkeypox is endemic to some regions} \textcolor{red}{ of Africa. Its primary symptoms, like those of other illnesses brought on by pox viruses, are fever and skin lesions.} \textcolor{red}{Unfortunately, a small percentage of individuals may experience severe, multi-systemic illness that frequently} requires hospitalization and, in some cases, intensive care unit support. \textcolor{red}{Due to} complexities in identifying \textcolor{red}{monkeypox virus}al strains and the ongoing evolution \textcolor{red}{of the} pandemic, effective treatments have yet \textcolor{red}{to be} approved. Antiviral drug treatment \textcolor{red}{of monkeypox} has primarily relied on small molecule therapies with established safety profiles for the treatment \textcolor{red}{of other} viral infections. These include the use of brincidofovir, a broad-spectrum DNA polyomavirus and poxvirus antiviral, and the nucleoside analogs, favipiravir and tecovirimat, which have shown potential in treating monkeypox in animal models. However, the efficacy \textcolor{red}{of these} drugs in clinical settings has yet \textcolor{red}{to be} demonstrated. In addition to the use of approved drugs, there is a pressing need to identify small molecules with specific activity against \textcolor{red}{monkeypox virus}. Natural product-derived small molecule libraries offer a rich source of potential leads with this type of specificity. In particular, compounds that inhibit poxvirus assembly and egress \textcolor{red}{have been} identified from the lanosterol library of Fasciola gigantica, a parasitic flatworm. The most potent of these, a compound designated Fas2\_32, was found \textcolor{red}{to have} 100-fold selectivity for the mammalian host over the sheep barber pole assembly sites used by the sheep infecting strain of Old World orthopoxvirus, vaccinia virus. In addition, Fas2\_32 was found \textcolor{red}{to be} active against the highly related \textcolor{red}{variola virus}, the causative agent \textcolor{red}{of smallpox}, with an EC50 of 1.3 03bcM. The epsilon toxin (ETX) \textcolor{red}{of monkeypox} is a 66 kDa envelope gly   \\ \hline

    \noalign{\hrule height 2pt} 
\end{xltabular}
}

\section{Limitations}
\label{sec:limitations}
While our results demonstrate that approximate unlearning in LLMs can leave recoverable traces in model parameters, our study is subject to several limitations.

First, our threat model assumes access to both pre- and post-unlearning model snapshots with white-box parameter access. While this setting is realistic in scenarios such as model auditing, rollback, or internal MLaaS pipelines where multiple checkpoints are maintained, it may not apply to strictly closed-source deployments where such access is restricted. 
Second, our reconstruction framework relies on parameter differences in the embedding layer to identify candidate tokens. Its effectiveness may vary under future unlearning techniques specifically designed to obscure such signals.
Finally, TRACE focuses on reconstructing textual data that has been explicitly targeted by the unlearning process. The reconstruction quality is influenced by factors such as model memorization strength, and unlearning batch size.

Despite these limitations, we believe our setting captures realistic deployment practices where multiple model versions are maintained, and our findings provide actionable insights for improving the safety and reliability of machine unlearning systems.



\end{document}

%% file: sections/1_introduction.tex
\section{Introduction}
The rapid adoption of large language models (LLMs) has intensified the need for effective data removal mechanisms, driven by regulatory requirements such as the EU’s GDPR~\cite{gdpr_art17} and AI Act~\cite{eu_ai_act_2024}, copyright compliance~\cite{zhang2025llms,hristov2016artificial,mueller2024llms},
and privacy~\cite{yan2025protecting,carlini2021extracting}.
Machine unlearning~\cite{bourtoule2021machine,nguyen2025survey,liu2025rethinking} aims to remove the influence of a specified subset of training data from a trained model. 
In practice, exact machine unlearning~\cite{bourtoule2021machine}, retraining models without the target data, remains computationally prohibitive for modern LLMs. 
As a result, mainstream real-world deployments rely on approximate unlearning methods~\cite{jang2023knowledge,feng2024fine,ilharco2022editing,pang2025label,liu2022continual,yao2024large,zhang2024negative}, which aim to forget specific data through targeted parameter updates without retraining. 
To support auditing, rollback, and safety verification, it is common to retain multiple model snapshots from before and after unlearning.

This raises a critical question: \emph{given access to original and unlearned models, to what extent do current unlearning methods leak sensitive information about the data they are intended to remove?}

Existing work~\cite{hu2024learn} shows that differences between the original and unlearned model parameters can be used to construct surrogate gradient signals, which are then exploited through gradient-matching inversion to reconstruct removed images.
However, this cannot be extended to generative LLMs, as the reconstruction optimization struggles to converge and often produces uninterpretable or disjointed outputs without the “anchor” provided by labels. 
More recently, efforts have been made to investigate the extraction of unlearned data from LLMs under the assumption that partial prefixes of the target texts are available, demonstrating that the remaining content can be recovered by exploiting discrepancies in output likelihoods between pre- and post-unlearning models~\cite{wu2025unlearned}. 

However, this reliance on prefix priors leaves an important open question: \emph{is unlearned data in LLMs secure without any data-specific prior knowledge?}

We answer this question by introducing TRACE, a framework for reconstructing text that has been intentionally unlearned by tracing latent parameter footprints between pre- and post-unlearning snapshots, without requiring any data-specific prior knowledge.
First, we identify that parameter differences in the embedding layer between pre- and post-unlearning models expose the tokens involved in the unlearning process, effectively reducing the reconstruction task from open-ended generation over the full vocabulary to assembling sequences from a compact set of candidate tokens.
Second, to resolve the combinatorial complexity of sequence reassembly, we develop a contrastive search strategy driven by a novel contrastive fluency score (CFS). This scoring mechanism integrates unlearning-induced divergence with linguistic plausibility, effectively navigating the search toward trajectories corresponding to unlearned sequences.
Third, to enable reconstruction of multiple unlearned samples, we design a two-stage “exploration-then-completion” architecture. Stage I performs broad exploration to capture diverse prefix candidates, while Stage II completes them into full sequences.
\noindent Our contributions are summarized as follows:
\begin{itemize}[wide=0pt, leftmargin=*, labelindent=0pt, label=$\bullet$]
    \item We reveal that approximate unlearning leaves a latent footprint in parameters, enabling reconstruction of text that has been intentionally unlearned, without any data-specific prior knowledge.
    \item We propose a framework named TRACE that achieves the reconstruction of multiple distinct unlearned samples without any data-specific prior knowledge, by utilizing parameter differences to effectively constrain the vast search space and employing a contrastive search strategy within a two-stage architecture.
    \item We conduct comprehensive experiments on various benchmarks including TOFU, MUSE, and WMDP, across multiple LLMs and six representative unlearning methods, showing that TRACE consistently achieves high-fidelity reconstruction. 
\end{itemize}

%% file: sections/2_related_work.tex
\section{Related Work}
\paragraph{Machine Unlearning in LLMs.}
Machine unlearning~\cite{bourtoule2021machine} aims to remove the influence of specific training samples from a trained model. 
Exact unlearning, which retrains the model on the remaining data, is widely regarded as the gold standard for data removal. 
However, exact unlearning is impractical for large language models (LLMs) due to their massive training scale and prohibitive computational cost. As a result, most existing work focuses on approximate unlearning via efficient parameter updates. 
Gradient ascent (GA)~\cite{jang2023knowledge,feng2024fine,ilharco2022editing,pang2025label} forgets samples by reversing learning on target data, i.e., applying updates that increase the loss on the data to be removed. Gradient difference
(GD)~\cite{liu2022continual} balances forgetting and utility by maximizing loss on forget data while minimizing loss on retain data. 
KL-regularized methods (KL)~\cite{yao2024large} introduce Kullback–Leibler divergence regularization on a retained set to preserve the model’s behavior on retained data during unlearning.
Negative Preference Optimization (NPO)~\cite{zhang2024negative} reformulates unlearning as preference optimization with a reference model to stabilize training.
Suppression-intended unlearning~\cite{ren2025sok,liu2024towards,ishibashi2023knowledge,xu2025suv,ji2024reversing,huang2024offset}
suppresses sensitive outputs rather than removing knowledge
and is beyond our scope.

\paragraph{Training Data Reconstruction.}
Recent studies show that training data can be reconstructed from models. Gradient-based approaches recover samples from leaked gradients~\cite{balunovic2022lamp,gupta2022recovering,petrov2024dager,xie2025recit}. To name a few, LAMP~\cite{balunovic2022lamp} reconstructs text using a language model prior and alternating optimization, while Gupta et al.~\cite{gupta2022recovering} use beam search with gradient-guided reordering. Petrov et al.~\cite{petrov2024dager} exploit low-rank self-attention gradients for exact inversion, and Xie et al.~\cite{xie2025recit} demonstrate reconstruction under parameter-efficient fine-tuning. Beyond gradients, reconstruction from internal representations has also been explored, such as embedding leakage~\cite{song2020information} and activation inversion attacks~\cite{dai2025stealing}.
However, these attacks rely on access to gradients or intermediate activations during training, and thus cannot be directly applied when only pre- and post-unlearning model snapshots are available.

\paragraph{Unlearning Inversion Attacks.}
Unlearning inversion attacks were first introduced in the context of image classification to reconstruct unlearned images by approximating the parameter differences between pre- and post-unlearning models as gradients of unlearned images~\cite{hu2024learn}.
Du et al.~\cite{du2024textual} extend this idea to textual classification, enabling the reconstruction of unlearned text. Ma et al. ~\cite{ma2026retrace} propose an RL-based approach that optimizes latent noise to reconstruct unlearned images. For LLMs, Wu et al.~\cite{wu2025unlearned} extract unlearned text in LLMs by leveraging memorization and probability gaps between original and unlearned models, assuming the adversary knows the identifiable prefix of the target unlearned text. 
However, optimizing-based methods ~\cite{hu2024learn,du2024textual,ma2026retrace} rely on labels, which are unavailable in generative LLMs, and prefix-based methods rely on the strong assumption of access to sample-specific context. Currently, no unlearning inversion approaches are available without prior information about unlearned data.

%% file: sections/3_threat_model.tex
\section{Threat Model}
\label{sec:threat_model}

\noindent\textbf{Scenario.}
Following prior unlearning inversion attacks~\cite{hu2024learn,du2024textual,ma2026retrace}, we consider practical model update workflows where both pre- and post-unlearning model versions are accessible. In Machine Learning as a Service (MLaaS) settings~\cite{hu2024learn}, historical model versions are often retained for rollback, auditing, debugging, or staged deployment, introducing a side channel, as discrepancies induced by unlearning may encode information about removed training data.   
A broader discussion of practical deployment scenarios is provided in Appendix~\ref{app:deployment}.

\noindent\textbf{Adversary.}
The adversary observes both pre- and post-unlearning models with white-box access to parameters, but has no access to the identities or content of the removed samples, the precise gradients or intermediate activations of the removed samples during training. The goal is to reconstruct samples removed through unlearning.

%% file: sections/4_attack.tex
\section{Method}
\label{sec:method}
\subsection{TRACE Overview}
\label{sec:Overview}

\begin{figure}[htbp]
  \centering
  \includegraphics[width=\linewidth]{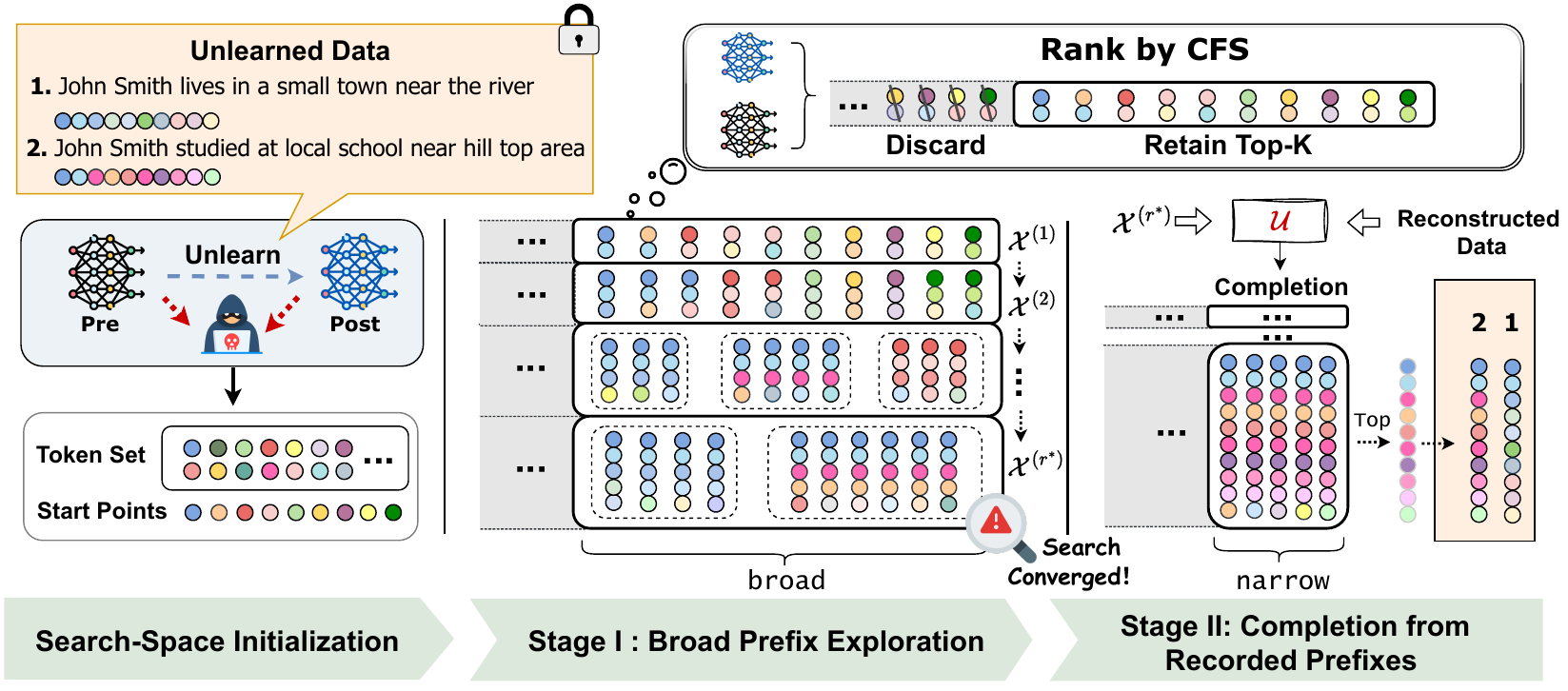}
    \caption{Framework of TRACE.}
    \label{fig:framework}
\end{figure}

TRACE reconstructs the unlearned data by incrementally assembling sequences from a compact set of candidate tokens, which are derived from embedding differences between the pre- and post-unlearning models via the search-space initialization (Sec.~\ref{sec:Initialization}).
Reconstruction follows an iterative search process, where candidate sequences are iteratively extended, scored by a novel contrastive fluency score (CFS)(Sec~\ref{sec:CFS}), and pruned to retain the highest-scoring candidates.
Building on the CFS-guided search, TRACE further structures the reconstruction process into two stages to facilitate the reconstruction of distinct unlearned samples (Sec~\ref{Sec:Two-Stage}): 
(1) \textbf{Stage I} - broad prefix exploration, which maintains a large set of candidate sequences at each step; 
(2) \textbf{Stage II} - completion from recorded prefixes, which completes the recorded prefixes from \textbf{Stage I} into full sequences with a relatively small candidate set at each step.
The framework of TRACE is shown in Fig.~\ref{fig:framework}.

\subsection{Search-Space Initialization}
\label{sec:Initialization}
\noindent\textbf{Search-Space Restriction via Candidate Tokens}.
\label{sec:candidate_token_extraction}
The large vocabulary of LLMs leads to a combinatorial explosion in possible sequences.
Inspired by prior findings that non-zero embedding gradients reveal tokens in training data~\cite{gupta2022recovering, melis2019exploiting}, we identify an analogous effect in unlearning: tokens involved in the unlearning process exhibit non-zero embedding parameter differences between the pre-unlearning model $\theta^*$ and post-unlearning model $\theta_u$. We exploit this property to construct a compact candidate token set to constrain the search space.
Let $E \in \mathbb{R}^{|\mathcal{V}| \times h}$ denote the input embedding matrix, 
where $|\mathcal{V}|$ is the vocabulary size and $h$ is the embedding dimension. 
For each token $v \in \mathcal{V}$, we compute the $\ell_2$ norm of the embedding difference $\Delta(v)$ and define the candidate token set $\mathcal{V}_f$ by:
\begin{equation}
\label{eq:candidate_token}
\mathcal{V}_f = \{ v \in \mathcal{V} \mid \Delta(v) > 0 \},
\quad
\Delta(v) = \left\| E_{\theta^*}(v) - E_{\theta_u}(v) \right\|_2.
\end{equation}

\noindent\textbf{Initialization of Starting Points.}
We construct a set of valid initial tokens $\mathcal{V}_0$ by filtering the candidate token set $\mathcal{V}_f$ based on linguistic constraints. 
Specifically, we retain those that are consistent with common sentence-initial patterns, such as tokens starting with uppercase letters or valid opening punctuation (e.g., \texttt{"}, \texttt{(}, \texttt{[}, \texttt{\{}, or \texttt{<}).

\subsection{Unified CFS-Guided Search}
\label{sec:CFS}
 TRACE follows an iterative search process, where candidate sequences are iteratively extended, scored using a novel contrastive fluency score (CFS), and pruned to retain the highest-scoring candidates for next step.
Let $\mathcal{X}^{(0)}$ denote the set of starting points, and $\mathcal{X}^{(r-1)}$ denote the retained set from step $r-1$.
We construct the extended candidate set at step $r$ by:
\begin{equation}
\label{eq:cr}
\mathcal{C}^{(r)} = \{\, x \oplus v \mid x \in \mathcal{X}^{(r-1)},\ v \in \mathcal{V}_f \,\},
\end{equation}
where $x \oplus v$ denotes the sequence obtained by extending $x$ with token $v$.
For an extended candidate sequence $x \in \mathcal{C}^{(r)}$, the perplexity~\cite{bahl1983maximum} under a model $\theta$ is given by:
\begin{equation}
\mathrm{PPL}_{\theta}(x)
=
\exp\!\left(
-\frac{1}{|x|}
\sum_{i=1}^{|x|}
\log p_{\theta}(x_i \mid x_{<i})
\right).
\end{equation}
Lower perplexity indicates that the sequence $x$ has higher likelihood. Unlearning reduces the likelihood of target data in the post-unlearning model, resulting in a divergence in perplexity between the pre- and post-unlearning models.
We leverage this to design a contrastive fluency score (CFS) to guide the search.
Specifically, for a candidate sequence $x \in \mathcal{C}^{(r)}$, the contrastive score is defined as:
\begin{equation}
\label{eq:contrastive_score}
\mathrm{CFS}(x)
=
\frac{\mathrm{PPL}_{\theta_u}(x)
-
\mathrm{PPL}_{\theta^*}(x)}
{\mathrm{PPL}_{\theta^*}(x)^{A(|x|)}},
\quad
A(|x|)=\alpha+\beta\log|x|.
\end{equation}
\begin{wrapfigure}{r}{0.5\linewidth}
\captionsetup[subfigure]{justification=centering,skip=-1pt}
\vspace{-8pt}
    \centering
    \begin{subfigure}[t]{0.49\linewidth}
        \centering
        \includegraphics[width=\linewidth]{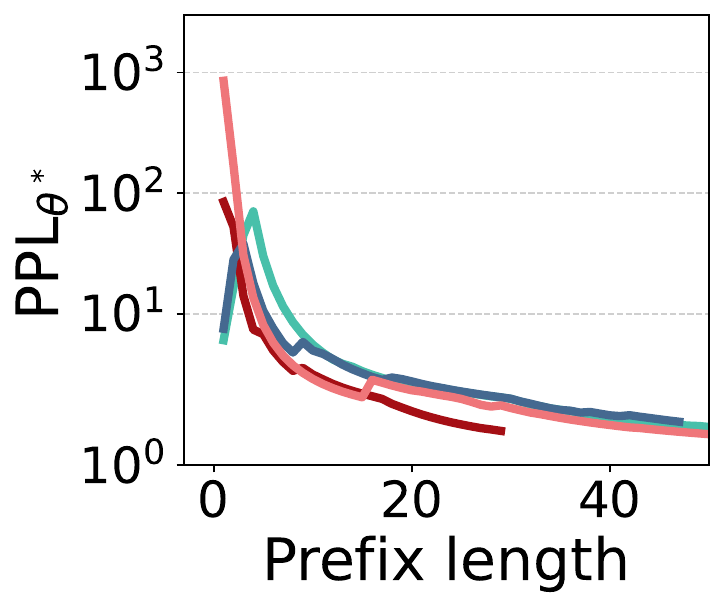}
        \caption{}
    \label{fig:ppl_pre_unlearning}
    \end{subfigure}
    \hfill
    \begin{subfigure}[t]{0.49\linewidth}
        \centering
        \includegraphics[width=\linewidth]{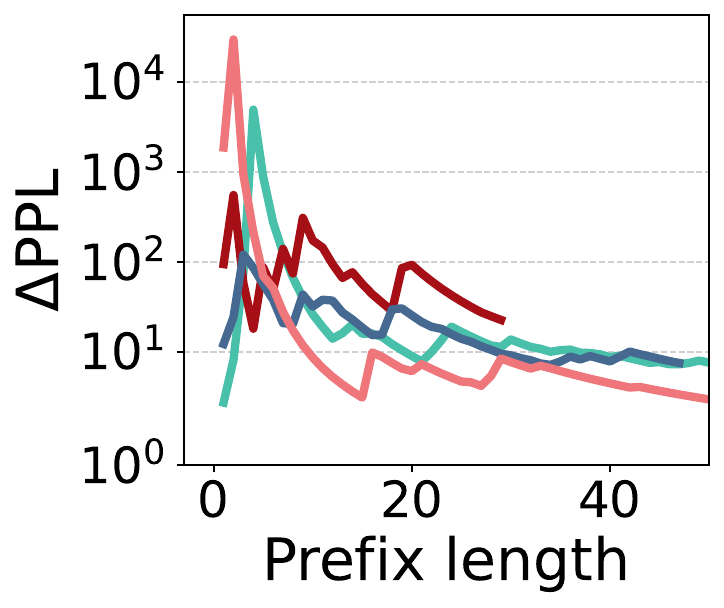}
        \caption{}
    \label{fig:delta_ppl}
    \end{subfigure}
    \caption{
    Effect of prefix length on perplexity for samples unlearned using GA.
    (a) Perplexity of unlearned samples under the pre-unlearning model.
    (b) Perplexity gap between pre- and post-unlearning models.
    }
    \label{fig:tokenwise_ppl}
\vspace{-10pt}
\end{wrapfigure}
The length-aware exponent $A(|x|)$ controls the relative importance of the fluency constraint $\mathrm{PPL_{\theta^*}}$, where $\alpha$ and $\beta$ are positive constants. 
This design is motivated by the fact that perplexity varies significantly with sequence length.
As shown in Figure~\ref{fig:ppl_pre_unlearning}, perplexity under the original model tends to be higher for shorter prefixes of unlearned data and decreases and stabilizes as the sequence grows.
By progressively increasing the influence of $\mathrm{PPL}_{\theta^*}(x)$, CFS remains tolerant of high-perplexity prefixes during early exploration when sufficiently strong contrastive signals (i.e., a large perplexity gap $\Delta \mathrm{PPL}$) are present. This prevents premature pruning of promising sequences, while enforcing stricter fluency constraints in later steps. At each step, candidates in $\mathcal{C}^{(r)}$ are ranked according to $\mathrm{CFS}(x)$, and a subset is retained to form $\mathcal{X}^{(r)}$, which serves as the input for the next iteration. The specific retention strategy is detailed in Sec.~\ref{Sec:Two-Stage}.

\subsection{Two-Stage Reconstruction via CFS-Guided Search}
\label{Sec:Two-Stage}
While CFS defines a scoring principle for candidate sequences, reconstructing multiple distinct unlearned samples remains challenging, as sequential search tends to concentrate on a number of highly similar high-scoring sequences. To address this issue, TRACE proposes a two-stage architecture that separates diversity exploration from sequence completion.
The procedure is summarized in Algorithm~\ref{alg:two_stage} in Appendix.

\noindent\textbf{Stage I: Broad Prefix Exploration}.
In this stage, TRACE aims to explore diverse candidate prefixes that may correspond to different unlearned samples. To this end, TRACE maintains a relatively large retained set at each step. 
As the unlearning signal $\Delta \mathrm{PPL}$ exhibits strong sample-to-sample variability, particularly for short prefixes (Fig.~\ref{fig:delta_ppl}), TRACE ranks candidates in $\mathcal{C}^{(r)}$ under two parameterizations of CFS:
$\mathrm{CFS}(\cdot;\alpha_s,\beta)$ and $\mathrm{CFS}(\cdot;\alpha_l,\beta),$
where $\alpha_s$ and $\alpha_l$ correspond to weaker and stronger fluency constraints, respectively, with $\alpha_s < \alpha_l$.
The former is more permissive to high-perplexity prefixes in early steps and thus preserves candidates with strong unlearning signals, while the latter favors more fluent candidates, even if their unlearning signals are relatively weak.
Accordingly, the retained set at step $r$ is given by:
\begin{equation}
\label{eq:xr}
\mathcal{X}^{(r)}
=
\operatorname{Top}_k\!\left(\mathcal{C}^{(r)}; \mathrm{CFS}(\cdot;\alpha_s,\beta)\right)
\cup
\operatorname{Top}_k\!\left(\mathcal{C}^{(r)}; \mathrm{CFS}(\cdot;\alpha_l,\beta)\right).
\end{equation}
To quantify diversity, sequences in $\mathcal{X}^{(r)}$ are grouped into clusters based on ROUGE-L similarity with a threshold $\tau$. Formally, $\mathcal{X}^{(r)}$ is partitioned into clusters $\mathcal{P}^{(r)}$ such that
\begin{equation}
\forall P \in \mathcal{P}^{(r)},\;
\forall x, x' \in P,\;
\mathrm{ROUGE\text{-}L}_{\text{recall}}(x, x') \ge \tau.
\label{eq:cluster}
\end{equation}
This stage continues until the retained set exhibits limited diversity, i.e., $|\mathcal{P}^{(r)}| \leq N$, 
where $N$ 
serves as an approximate indication of how many distinct samples the attacker aims to reconstruct.
The last retained set is recorded as $\mathcal{X}^{(r^*)}$, serving as candidate prefixes, which are then completed in Stage II. 
We use a broad search width $k = N \times d$ in stage I, where $d$ is an expansion factor, to increase the chance of retaining prefixes corresponding to true unlearned samples.

\noindent\textbf{Stage II: Completion from Recorded Prefixes.}
This stage revisits the last recorded prefix set $\mathcal{X}^{(r^*)}$ from Stage I and completes it into full sequences via a narrower CFS-guided search, where the strategy of extending and pruning candidates at each step follows Eq.~\eqref{eq:cr} and Eq.~\eqref{eq:xr} but with a smaller search width $k = Q$. 
To avoid redundant reconstruction, let $\mathcal{S}$ denote the set of reconstructed sequences. The set of uncovered prefixes $\mathcal{U}$ is:
\begin{equation}
\label{eq:revisist}
\mathcal{U} = \left\{ x \in \mathcal{X}^{(r^*)} \;\middle|\; \max_{s \in \mathcal{S}} \mathrm{ROUGE\text{-}L}_{\text{recall}}(x, s) < \tau \right\}.
\end{equation}
TRACE performs iterative CFS guided search over \(\mathcal{U}\). In each iteration, the search continues until a termination condition is met, yielding newly reconstructed sequences and triggering corresponding updates to \(\mathcal{S}\) and \(\mathcal{U}\).
This process repeats until $\mathcal{U} = \emptyset$, i.e., all prefixes in $\mathcal{X}^{(r^*)}$ are covered. Detailed stopping criteria are provided in Appendix~\ref{sec:stop}.

%% file: sections/5_experiment.tex
\section{Experiments}
\label{sec:exp}
\subsection{Setup}
\label{sec:setup}
We evaluate TRACE on various standard unlearning benchmarks, including TOFU~\cite{maini2024tofu}, MUSE~\cite{shi2024muse}, and WMDP~\cite{li2024wmdp}. 
For each dataset, we first fine-tune the model to obtain a pre-unlearning checkpoint, and then apply six representative unlearning methods, including gradient ascent (GA), gradient difference (GD), KL minimization (KL), negative preference optimization (NPO), NPO-Diff~\cite{maini2024tofu}, and NPO-KL~\cite{maini2024tofu}, to unlearn the target samples and obtain the corresponding unlearned models.
Experiments are conducted on multiple open-source LLMs, including  LLaMA2-13B, LLaMA2-7B~\cite{touvron2023llama},Mistral-7B~\cite{jiang2023mistral7b}, Qwen2.5-3B~\cite{bai2023qwen}, StableLM-3B-4E1T~\cite{tow2023stablelm} and TinyLlama-1.1B~\cite{zhang2024tinyllama}.
Follow~\cite{wu2025unlearned}, three metrics are adopted for evaluation, including ROUGE-L recall (R), and average extraction success rate (A-ESR$_\kappa$), 
which measures the percentage of ground-truth unlearned samples that can be matched by reconstructed samples under the ROUGE-L recall threshold $\kappa \in \{1.0, 0.9\}$ (as formally defined in Appendix~\ref{sec:metric}).
We set $N=1.5\times B$ in our experiments, where $B$ is the
unlearning batch size. TRACE also involves the CFS coefficients
$(\alpha_s,\alpha_l,\beta)$, similarity threshold $\tau$, and search
parameters $d$ and $Q$. We analyze their sensitivity in
Sec.~\ref{sec:ablation}; complete experimental settings and detailed
explanations of their roles are provided in
Appendices~\ref{exp_details} and~\ref{app:hyperparameters}, respectively.
\subsection{Multi-sample reconstruction without prior knowledge of unlearned data}

\begin{table*}[t]
\centering
\caption{Reconstruction performance on TOFU and MUSE.}
\label{tab:unlearning_results}

\renewcommand{\arraystretch}{1.3} 
\setlength{\aboverulesep}{0pt}
\setlength{\belowrulesep}{0pt}
\setlength{\tabcolsep}{4pt} 

\scalebox{0.59}{
    \begin{NiceTabular}{ccc cccccccc cccc}
    \toprule
    \rowcolor{headergray} 
    & & & \multicolumn{2}{c}{GA} & \multicolumn{2}{c}{GD} & 
    \multicolumn{2}{c}{KL} & \multicolumn{2}{c}{NPO} & 
    \multicolumn{2}{c}{NPO-Diff} & \multicolumn{2}{c}{NPO-KL} \\
    
    \cmidrule{1-15} 
    
    \rowcolor{headergray} 
    Dataset & Model & Metric 
    & Ours & PPL-only & Ours & PPL-only & Ours & PPL-only & Ours & PPL-only & Ours & PPL-only & Ours & PPL-only \\
    \midrule
    
    \multirow{15}{*}{\rotatebox{90}{\textbf{TOFU}}} 

    & \multirow{3}{*}{TinyLlama-1.1B}
    & ROUGE-L(R) & \cellcolor{oursbg}0.9084 & 0.4383 & \cellcolor{oursbg}0.8770  & 0.1770 & \cellcolor{oursbg}0.8785 & 0.1770 & \cellcolor{oursbg}0.8882
 & 0.4383 & \cellcolor{oursbg}0.8358 & 0.1770 & \cellcolor{oursbg}0.8472 & 0.1770 \\
    & & A-ESR$_{1.0}$(\%) & \cellcolor{oursbg}78.75 &10.00 & \cellcolor{oursbg}71.25 & 2.50 & \cellcolor{oursbg}76.25 & 2.50 & \cellcolor{oursbg}76.25
 & 10.00 & \cellcolor{oursbg}68.75 & 2.50 & \cellcolor{oursbg}67.50 & 2.50 \\
    & & A-ESR$_{0.9}$(\%) & \cellcolor{oursbg}86.25 & 15.00 & \cellcolor{oursbg}82.50 & 2.50 & \cellcolor{oursbg}82.50 & 2.50 & \cellcolor{oursbg}82.50
 & 15.00 & \cellcolor{oursbg}76.25 & 2.50 & \cellcolor{oursbg}75.00 & 2.50 \\
    \cmidrule{2-15}

    & \multirow{3}{*}{Qwen2.5-3B}
    & ROUGE-L(R) & \cellcolor{oursbg}0.8409 & 0.1681 & \cellcolor{oursbg}0.8191 & 0.1387 & \cellcolor{oursbg}0.8362 & 0.1386 & \cellcolor{oursbg}0.8189
 & 0.1592 & \cellcolor{oursbg}0.8259 & 0.1534 & \cellcolor{oursbg}0.8625 & 0.1586 \\
    & & A-ESR$_{1.0}$(\%) & \cellcolor{oursbg}67.50 & 0.00 & \cellcolor{oursbg}68.75 & 0.00 & \cellcolor{oursbg}66.25 & 0.00 & \cellcolor{oursbg}65.00
 & 0.00 & \cellcolor{oursbg}67.50 & 0.00 & \cellcolor{oursbg}71.25 & 0.00 \\
    & & A-ESR$_{0.9}$(\%) & \cellcolor{oursbg}76.25 & 1.25 & \cellcolor{oursbg}75.00 & 1.25 & \cellcolor{oursbg}76.25 & 1.25 & \cellcolor{oursbg}75.00
 & 1.25 & \cellcolor{oursbg}76.25 & 1.25 & \cellcolor{oursbg}78.75 & 1.25 \\
    \cmidrule{2-15}
    
    & \multirow{3}{*}{LLaMA2-7B}
    & ROUGE-L(R) & \cellcolor{oursbg}0.9799 & 0.6017 & \cellcolor{oursbg}0.9695 & 0.2946 & \cellcolor{oursbg}0.9592 & 0.2946 & \cellcolor{oursbg}0.9715 & 0.6017 & \cellcolor{oursbg}0.9578 & 0.2946 & \cellcolor{oursbg}0.9684 & 0.2946 \\
    & & A-ESR$_{1.0}$(\%) & \cellcolor{oursbg}95.00 & 28.75 & \cellcolor{oursbg}92.50 & 7.50 & \cellcolor{oursbg}90.00 & 7.50 & \cellcolor{oursbg}93.75 & 28.75 & \cellcolor{oursbg}93.75 & 7.50 & \cellcolor{oursbg}96.25 & 7.50 \\
    & & A-ESR$_{0.9}$(\%) & \cellcolor{oursbg}97.50 & 36.25 & \cellcolor{oursbg}96.25 & 10.00 & \cellcolor{oursbg}91.25 & 10.00 & \cellcolor{oursbg}95.00 & 36.25 & \cellcolor{oursbg}95.00 & 10.00 & \cellcolor{oursbg}96.25 & 10.00 \\

    \cmidrule{2-15}
    & \multirow{3}{*}{Mistral-7B}
    & ROUGE-L(R) & \cellcolor{oursbg}0.9594 &  0.7145 & \cellcolor{oursbg}0.9873 & 0.5584 & \cellcolor{oursbg} 0.8707 & 0.5584 & \cellcolor{oursbg} 0.9602 & 0.7145 & \cellcolor{oursbg}0.9604 & 0.5584 & \cellcolor{oursbg} 0.9299 & 0.5584 \\
    & & A-ESR$_{1.0}$(\%) & \cellcolor{oursbg}90.00 & 47.50 & \cellcolor{oursbg}92.50 & 37.50 & \cellcolor{oursbg} 61.25 & 37.50 & \cellcolor{oursbg}90.00 & 47.50 & \cellcolor{oursbg}88.75 & 37.50 & \cellcolor{oursbg}82.50 & 37.50 \\
    & & A-ESR$_{0.9}$(\%) & \cellcolor{oursbg}95.00 & 51.25 & \cellcolor{oursbg}98.75 & 37.50 & \cellcolor{oursbg}67.50 & 37.50 & \cellcolor{oursbg}95.00 & 51.25 & \cellcolor{oursbg}93.75 & 37.50 & \cellcolor{oursbg}86.25 & 37.50 \\
    
    \cmidrule{2-15}
    & \multirow{3}{*}{LLaMA2-13B}
    & ROUGE-L(R) & \cellcolor{oursbg}0.9936 & 0.6106 & \cellcolor{oursbg}0.9792 & 0.3311 & \cellcolor{oursbg}0.9792 & 0.3311 & \cellcolor{oursbg}0.9795 & 0.6106 & \cellcolor{oursbg}0.9703 & 0.3311 & \cellcolor{oursbg}0.9690 & 0.3311 \\
    & & A-ESR$_{1.0}$(\%) & \cellcolor{oursbg}93.75 & 33.75 & \cellcolor{oursbg}92.50 & 10.00 & \cellcolor{oursbg}93.75 & 10.00 & \cellcolor{oursbg}96.25 & 33.75 & \cellcolor{oursbg}93.75 & 10.00 & \cellcolor{oursbg}93.75 & 10.00 \\
    & & A-ESR$_{0.9}$(\%) & \cellcolor{oursbg}98.75 & 40.00 & \cellcolor{oursbg}97.50 & 10.00 & \cellcolor{oursbg}97.50 & 10.00 & \cellcolor{oursbg}96.25 & 40.00 & \cellcolor{oursbg}95.00 & 10.00 & \cellcolor{oursbg}95.00 & 10.00 \\
    
    \midrule
    \midrule
    
    \multirow{15}{*}{\rotatebox{90}{\textbf{MUSE}}} 
    
    & \multirow{3}{*}{TinyLlama-1.1B}
        & ROUGE-L(R) & \cellcolor{oursbg}0.8750 & 0.2834 & \cellcolor{oursbg}0.7555 & 0.1153 & \cellcolor{oursbg}0.7641 & 0.1153 & \cellcolor{oursbg}0.8467
     & 0.2834 & \cellcolor{oursbg}0.8136 & 0.1153 & \cellcolor{oursbg}0.8099 & 0.1153 \\
        & & A-ESR$_{1.0}$(\%) & \cellcolor{oursbg}80.00 & 17.50 & \cellcolor{oursbg}65.00 & 3.75 & \cellcolor{oursbg}67.50 & 3.75 & \cellcolor{oursbg}76.25
     & 17.50 & \cellcolor{oursbg}71.25 &3.75 & \cellcolor{oursbg}76.25 & 3.75 \\
        & & A-ESR$_{0.9}$(\%) & \cellcolor{oursbg}83.75 & 20.00 & \cellcolor{oursbg}67.50 & 5.00 & \cellcolor{oursbg}70.00 & 5.00 & \cellcolor{oursbg}78.75
     & 20.00 & \cellcolor{oursbg}75.00 & 5.00 & \cellcolor{oursbg}77.50 & 5.00 \\
        \cmidrule{2-15}
    
    & \multirow{3}{*}{Stablelm-3B}
    & ROUGE-L(R) & \cellcolor{oursbg}0.8633 & 0.5624 & \cellcolor{oursbg}0.7448 & 0.4246 & \cellcolor{oursbg}0.7883 & 0.4246 & \cellcolor{oursbg}0.8838 & 0.5624 & \cellcolor{oursbg}0.7958 & 0.4246 & \cellcolor{oursbg}0.7728 & 0.4246 \\
    & & A-ESR$_{1.0}$(\%) & \cellcolor{oursbg}80.00 & 46.25 & \cellcolor{oursbg}62.50 & 32.50 & \cellcolor{oursbg}72.50 & 32.50 & \cellcolor{oursbg}83.75 & 46.25 & \cellcolor{oursbg}70.00 & 32.50 & \cellcolor{oursbg}68.75 & 32.50 \\
    & & A-ESR$_{0.9}$(\%) & \cellcolor{oursbg}85.00 & 51.25 & \cellcolor{oursbg}71.25 & 36.25 & \cellcolor{oursbg}78.75 & 36.25 & \cellcolor{oursbg}87.50 & 51.25 & \cellcolor{oursbg}76.25 & 36.25 & \cellcolor{oursbg}75.00 & 36.25 \\
    
    \cmidrule{2-15}
    & \multirow{3}{*}{LLaMA2-7B}
    & ROUGE-L(R) & \cellcolor{oursbg}0.9457 &  0.3915 & \cellcolor{oursbg}0.9332 & 0.1654 & \cellcolor{oursbg}0.9398 & 0.1654 & \cellcolor{oursbg}0.9579 & 0.3915 & \cellcolor{oursbg}0.9554 & 0.1654 & \cellcolor{oursbg}0.9646 & 0.1654 \\
    & & A-ESR$_{1.0}$(\%) & \cellcolor{oursbg}93.75 & 37.50 & \cellcolor{oursbg}92.50 & 13.75 & \cellcolor{oursbg}93.75 & 13.75 & \cellcolor{oursbg}95.00 & 37.50 & \cellcolor{oursbg}93.75 & 13.75 & \cellcolor{oursbg}96.25 & 13.75 \\
    & & A-ESR$_{0.9}$(\%) & \cellcolor{oursbg}93.75 & 37.50 & \cellcolor{oursbg}92.50 & 13.75 & \cellcolor{oursbg}93.75 & 13.75 & \cellcolor{oursbg}95.00 & 37.50 & \cellcolor{oursbg}93.75 & 13.75 & \cellcolor{oursbg}96.25 & 13.75 \\

    \cmidrule{2-15}
    & \multirow{3}{*}{Mistral-7B}
    & ROUGE-L(R) & \cellcolor{oursbg}0.9549 & 0.3533 & \cellcolor{oursbg}0.8753 & 0.1221 & \cellcolor{oursbg}0.9760 &  0.1221 & \cellcolor{oursbg}0.9637 & 0.3533 & \cellcolor{oursbg}0.9540 & 0.1221 & \cellcolor{oursbg}0.9604 & 0.1221  \\
    & & A-ESR$_{1.0}$(\%) & \cellcolor{oursbg}93.75 & 31.25 & \cellcolor{oursbg}83.75 & 6.25 & \cellcolor{oursbg}96.25 & 6.25 & \cellcolor{oursbg}94.44 & 31.25 & \cellcolor{oursbg}90.00 & 6.25 & \cellcolor{oursbg}93.75 &  6.25 \\
    & & A-ESR$_{0.9}$(\%) & \cellcolor{oursbg}93.75 & 32.50 & \cellcolor{oursbg}85.00 & 7.50 & \cellcolor{oursbg}96.25 & 7.50 & \cellcolor{oursbg}94.44 & 32.50 & \cellcolor{oursbg}93.75 & 7.50 & \cellcolor{oursbg}93.75 & 7.50 \\
    
    \cmidrule{2-15}
    & \multirow{3}{*}{LLaMA2-13B}
    & ROUGE-L(R) & \cellcolor{oursbg}0.9327 & 0.3839 & \cellcolor{oursbg}0.9438 & 0.1728 & \cellcolor{oursbg}0.9628 & 0.1728 & \cellcolor{oursbg}0.9441 & 0.3839 & \cellcolor{oursbg}0.9399 & 0.1728 & \cellcolor{oursbg}0.9513 & 0.1728 \\
    & & A-ESR$_{1.0}$(\%) & \cellcolor{oursbg}92.50 & 35.00 & \cellcolor{oursbg}92.50 & 12.50 & \cellcolor{oursbg}96.25 & 12.50 & \cellcolor{oursbg}92.50 & 35.00 & \cellcolor{oursbg}92.50 & 12.50 & \cellcolor{oursbg}95.00 & 12.50 \\
    & & A-ESR$_{0.9}$(\%) & \cellcolor{oursbg}92.50 & 36.25 & \cellcolor{oursbg}92.50 & 13.75 & \cellcolor{oursbg}96.25 & 13.75 & \cellcolor{oursbg}92.50 & 36.25 & \cellcolor{oursbg}92.50 & 13.75 & \cellcolor{oursbg}95.00 & 13.75 \\
    \bottomrule
    \end{NiceTabular}
    }
\vspace{-10pt}
\label{tab:main}
\end{table*}

We evaluate TRACE against a PPL-only baseline that adopts the same framework as TRACE but ranks candidates solely by perplexity under the pre-unlearning model. This baseline serves as an ablation to isolate the contribution of the CFS mechanism from the original model's inherent memorization. In
Appendix~\ref{app:baseline_discussion}, we further compare
PPL-only with FILM~\cite{gupta2022recovering}, 
and show that PPL-only is a
stronger reconstruction baseline.
Table~\ref{tab:main} summarizes the results averaged over 10 randomly sampled batches with an unlearning size of $B=8$. TRACE consistently achieves superior performance across various models and unlearning methods. 
While the efficacy of PPL-only deteriorates significantly under unlearning methods that incorporate retain-set regularization (e.g., GD, KL, NPO-Diff/KL), TRACE remains remarkably robust across all such settings.
Furthermore, we observe that larger models are more susceptible to reconstruction than smaller ones due to their superior capacity for memorization. For instance, TRACE achieves ROUGE-L(R) above 0.93 on LLaMA2-7B, Mistral-7B, and LLaMA2-13B. We provide a visualization of reconstructed data in Appendix~\ref{sec:multi_sample_recon}.


\begin{wraptable}{r}{0.63\linewidth}
\vspace{-8pt}
\caption{TRACE Performance under different unlearning batch sizes using LLaMA2-7B.}
\label{tab:batch_size_comparison}
\small
\renewcommand{\arraystretch}{1.3} 
\setlength{\aboverulesep}{0pt}
\setlength{\belowrulesep}{0pt}
\setlength{\tabcolsep}{4pt} 

\scalebox{0.62}{
    
    \begin{tabular}{cc cccc  cccc}
    \toprule
    &  & \multicolumn{4}{c}{GA} & \multicolumn{4}{c}{NPO} \\
    \cmidrule(lr){1-10} 
    Dataset & Metric & $B=8$ & $B=16$ & $B=32$ & $B=128$ & $B=8$ & $B=16$ & $B=32$ & $B=128$\\
    \midrule
    
    \multirow{3}{*}{\rotatebox{90}{\textbf{TOFU}}} 
    & ROUGE-L(R) & 0.9799 & 0.9827 & 0.9541  &  0.7924 & 0.9715 & 0.9569 & 0.9287 & 0.8146\\
    & A-ESR$_{1.0}$(\%) & 95.00 & 92.50 & 88.75& 71.09 & 93.75 & 87.50 & 87.50& 73.44 \\
    & A-ESR$_{0.9}$(\%) & 97.50 & 97.50 & 92.50& 74.22 & 95.00 & 92.50 & 90.62& 75.78 \\
    
    \midrule
    
    \multirow{3}{*}{\rotatebox{90}{\textbf{MUSE}}} 
    & ROUGE-L(R) & 0.9457 & 0.9408 & 0.8996& 0.8007 & 0.9579 & 0.9561 & 0.8897& 0.8051 \\
    & A-ESR$_{1.0}$(\%)  & 93.75 & 92.50 & 87.81 & 74.22& 95.00 & 93.75 & 86.56 & 74.22\\
    & A-ESR$_{0.9}$(\%)  & 93.75 & 92.50 & 88.12& 77.34 & 95.00 & 94.38 & 87.18 & 81.25\\
    
    \bottomrule
    \end{tabular}
}
\label{tab:batch}
\vspace{-5pt}
\end{wraptable}
Table~\ref{tab:batch} further evaluates different unlearning batch sizes $B \in \{8,16,32,128\}$. TRACE remains effective even at $B=128$, achieving ROUGE-L(R) scores above 0.8 on MUSE. Although performance decreases as batch size increases, this degradation can be mitigated by increasing expansion factor $d$, which allows more candidate sequences to be retained per step and improves the chance of capturing true prefixes corresponding to unlearned samples (see Section~\ref{sec:quota_d}).

\subsection{Comparison with Prefix-Assisted Baseline UBNF~\cite{wu2025unlearned}}
\label{sec:wmdp_long}
\begin{figure}[htbp]
  \centering
  \includegraphics[width=0.65\linewidth]{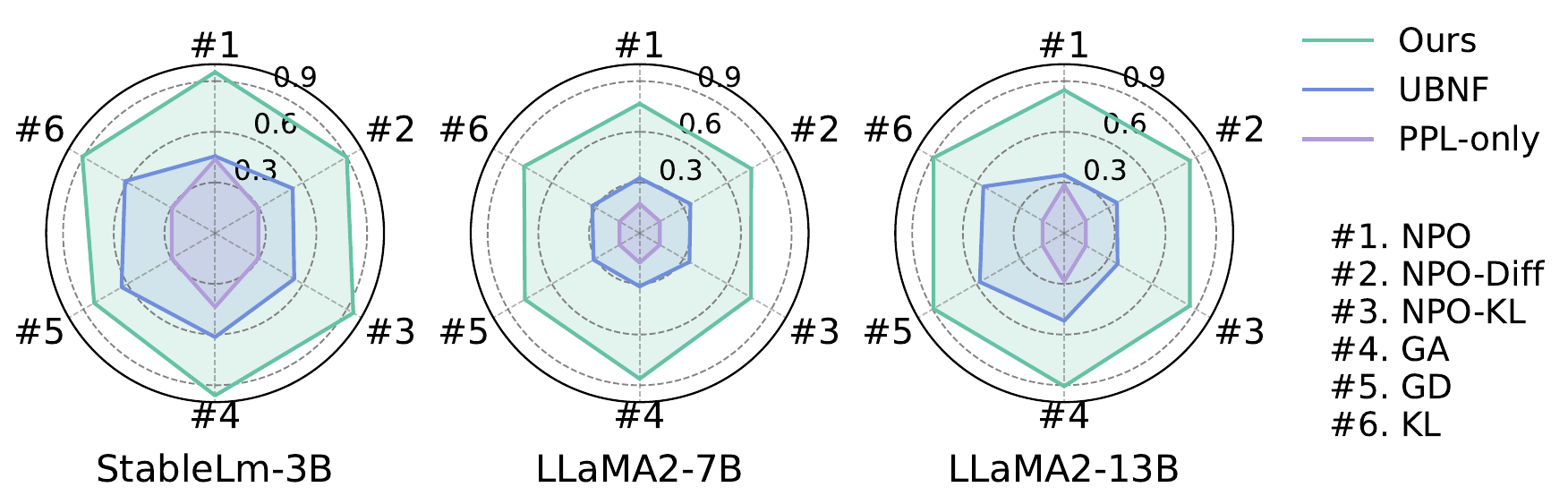}
    \caption{Long-form reconstruction performance on WMDP measured by ROUGE-L(R).}
    \label{fig:WMDP}
    \vspace{-5pt}
\end{figure}

To enable a fair comparison with the prefix-conditioned unlearned text reconstruction setting in UBNF~\cite{wu2025unlearned},
we conduct reconstruction on the long-form WMDP dataset, where the goal is to reconstruct a long sensitive text. Each WMDP sample is a 512-token text, and UBNF~\cite{wu2025unlearned} is given the first 10 words as a prefix, while TRACE uses none.
As shown in Fig.~\ref{fig:WMDP}, TRACE consistently outperforms both baselines across models and six unlearning methods without the strong assumptions in the known prefix. In particular, on LLaMA2-7B, TRACE improves ROUGE-L(R) by more than 100\% over UBNF. These results demonstrate the effectiveness of TRACE in realistic long-form reconstruction scenarios. We provide a visualization of reconstructed text in Appendix~\ref{sec:example_WMDP}.

\paragraph{Study Case: Prefix-Matched Reconstruction on RWKU Dataset~\cite{cao2024rwku}.}

\begin{wrapfigure}{r}{0.6\linewidth}
\vspace{-10pt}
  \centering
\includegraphics[width=0.99\linewidth]{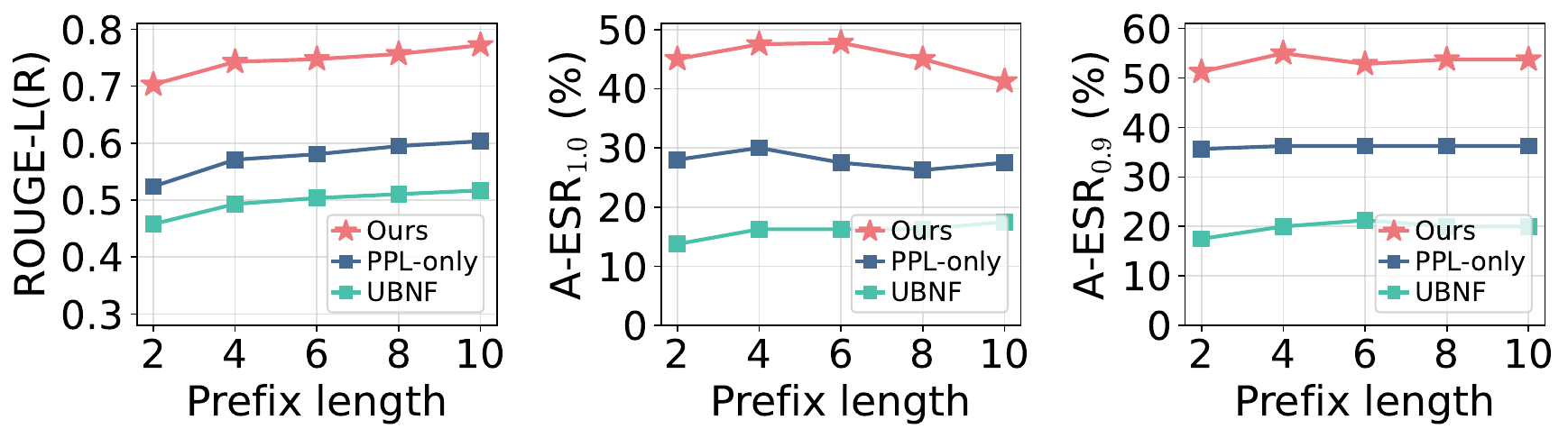}
    \caption{Reconstruction performance with prefix on RWKU using LLaMA2-7B}
    \label{fig:prefixlen_compare}
\vspace{-10pt}
\end{wrapfigure}
We further evaluate prefix-assisted reconstruction on the RWKU dataset, where unlearned samples have an average length of 150 words. We compare three methods given equal-length prefixes (2–10 words) from the unlearned data, using GA with an unlearning batch size of 16.
As shown in Fig.~\ref{fig:prefixlen_compare}, TRACE consistently outperforms UBNF, demonstrating its effectiveness under limited prefix information in recovering the remaining content.

\subsection{Ablation Study}
\label{sec:ablation}

\paragraph{Hyperparameters $\alpha_s$, $\alpha_l$ and $\beta$ in CFS.}
\begin{wrapfigure}{r}{0.6\linewidth}
\vspace{-10pt}
  \centering
\includegraphics[width=0.99\linewidth]{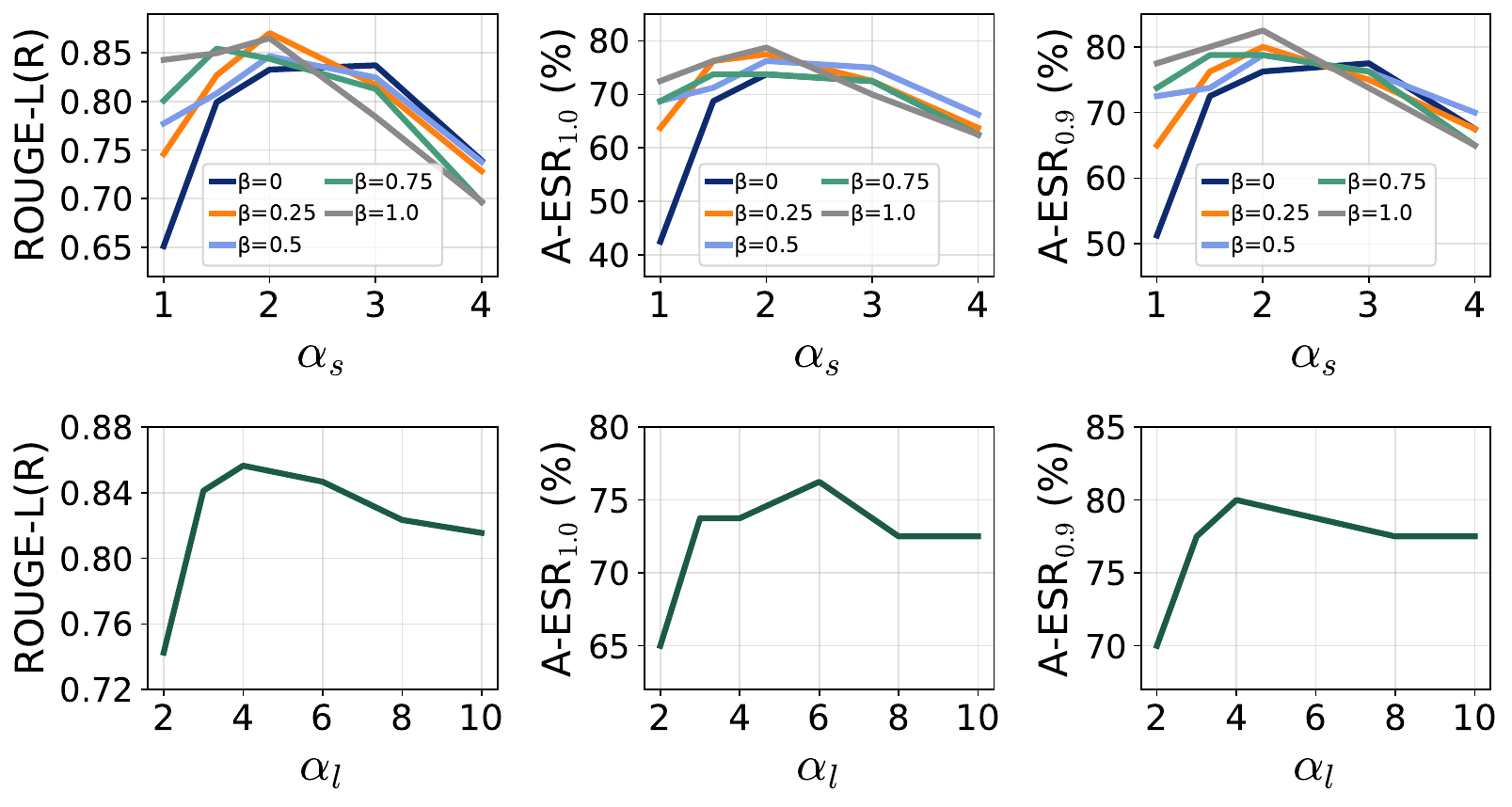}
    \caption{Hyperparameters in $\mathrm{CFS}$ on MUSE. Top: varying $\alpha_s$ and $\beta$ ($\alpha_l = 6.0$). Bottom: varying $\alpha_l$ ($\alpha_s=2, \beta=0.5$).}
    \label{fig:ablation_alpha_beta}
\vspace{-10pt}
\end{wrapfigure}
Figure~\ref{fig:ablation_alpha_beta} shows the sensitivity of CFS to $\alpha_s$, $\alpha_l$ and $\beta$ on MUSE using TinyLlama under NPO. 
Introducing the length-aware term $\beta > 0$ improves the robustness of CFS, preventing performance collapse at small $\alpha_s$ 
, with the best results around $\alpha_s=2$. 
This suggests a beneficial dynamic trade-off: early steps prioritize strong unlearning signals even if prefixes are not fluent, while later steps increasingly favor linguistic plausibility as sequences become more complete.
Regarding $\alpha_l$, setting $\alpha_l = \alpha_s = 2$ degrades performance as the dual-ranking reverts to a single objective. 
$\alpha_l \in [3, 8]$ leads to stronger performance, 
suggesting that this branch serves as a complement to the small $\alpha_s$ setting by capturing fluent prefixes with weaker unlearning signals that may otherwise be overlooked.



\paragraph{Parameter $d$ in stage I.}
\label{sec:quota_d}

\begin{wraptable}{r}{0.35\linewidth}
\vspace{-3pt}
\centering
\caption{Effect of $d$ on MUSE.}
\small
\label{tab:quota_exploration}
\renewcommand{\arraystretch}{1.3} 
\setlength{\aboverulesep}{0pt}
\setlength{\belowrulesep}{0pt}
\setlength{\tabcolsep}{4pt} 
\scalebox{0.8}{
    \begin{tabular}{l ccc ccc}
    \toprule
    Metric & $d=5$ & $d=10$ & $d=15$ \\
    \midrule
    
    ROUGE-L(R) & 0.8897 & 0.9504 & 0.9564 \\
    A-ESR$_{1.0}$(\%)  & 86.56 & 92.50 & 92.81 \\
    A-ESR$_{0.9}$(\%)  & 87.19 & 93.13 & 93.75 \\
    
    \bottomrule
    \end{tabular}
}
\label{tab:quota}
\vspace{-8pt}
\end{wraptable} 
The parameter $d$ determines how many candidates are retained at each step in stage I and therefore directly affects the ability to capture diverse prefixes corresponding to different unlearned samples.
As observed in Table~\ref{tab:batch}, increasing the unlearning batch size $B$ to 32 (with a fixed $d=5$) leads to performance degradation on the MUSE dataset. 
Table~\ref{tab:quota} reports that increasing $d$ improves reconstruction performance across metrics, indicating that TRACE can scale to larger batch unlearning settings by increasing $d$.

\paragraph{Parameter $Q$ in stage II.}
\begin{wraptable}{r}{0.35\linewidth}
\vspace{-10pt}
\centering
\caption{Effect of $Q$ on MUSE.}
\small
\label{tab:quota_exploration}
\renewcommand{\arraystretch}{1.3} 
\setlength{\aboverulesep}{0pt}
\setlength{\belowrulesep}{0pt}
\setlength{\tabcolsep}{4pt} 
\scalebox{0.8}{
    \begin{tabular}{l ccc ccc}
    \toprule
    Metric & $Q=1$ & $Q=5$ & $Q=10$ \\
    \midrule
    
    ROUGE-L(R) & 0.8148 & 0.8529 & 0.8467 \\
    A-ESR$_{1.0}$(\%)  & 68.75 & 76.25 & 76.25 \\
    A-ESR$_{0.9}$(\%)  & 76.25 & 78.75 & 77.50 \\
    
    \bottomrule
    \end{tabular}
}
\label{tab:quota_Q2}
\vspace{-8pt}
\end{wraptable}
The parameter $Q$ controls the search width in Stage II. As the search focuses on extending selected prefixes rather than exploring diverse trajectories, a narrow search width is sufficient.
As shown in Table~\ref{tab:quota_Q2}, a relatively small value (e.g., $Q=5$) achieves strong performance.
These results highlight the effectiveness of decoupling diversity discovery from sequence completion, enabling Stage II to achieve reliable reconstruction with a modest search budget while avoiding redundant exploration.


\paragraph{Effect of Similarity Threshold $\tau$.}
\begin{wrapfigure}{r}{0.6\linewidth}
\vspace{-8pt}
  \centering
\includegraphics[width=0.99\linewidth]{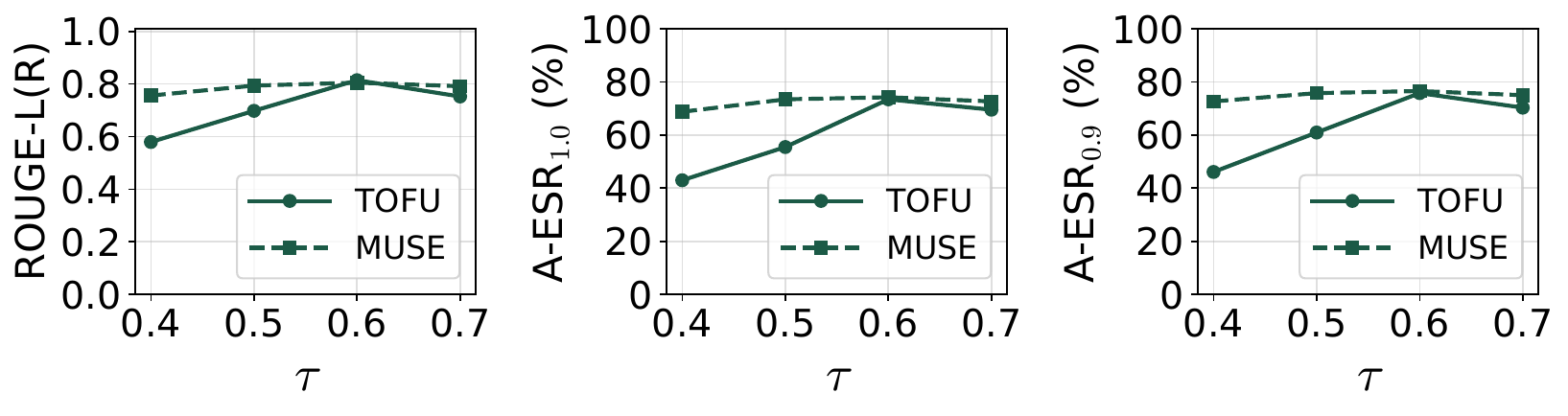}
    \caption{Effect of Similarity Threshold $\tau$}
    \label{fig:rouge}
    \vspace{-5pt}
\end{wrapfigure}
We study the impact of the similarity threshold $\tau$ used in Eq.~\eqref{eq:cluster} and Eq.~\eqref{eq:revisist}, using LLaMA2-7B under NPO with unlearning batch size $B=128$. As shown in Fig.~\ref{fig:rouge}, 
Reconstruction performance on TOFU demonstrates a higher sensitivity to the choice of $\tau$ compared to that on MUSE.
We attribute this to its structural properties: TOFU consists of question--answer pairs, where samples tend to share higher lexical and semantic similarity. When $\tau$ is too small, structurally similar but semantically different prefixes are merged into the same cluster, leading to inadequate reconstruction of distinct samples.
Empirically, setting $\tau \in [0.6, 0.7]$ is found to provide a robust balance between distinguishing distinct sequences and merging redundant sequences.

\paragraph{Training and Unlearning Epochs.}
\label{sec:ablation_epoch}
\begin{wrapfigure}{r}{0.6\linewidth}
\vspace{-10pt}
  \centering
\includegraphics[width=0.99\linewidth]{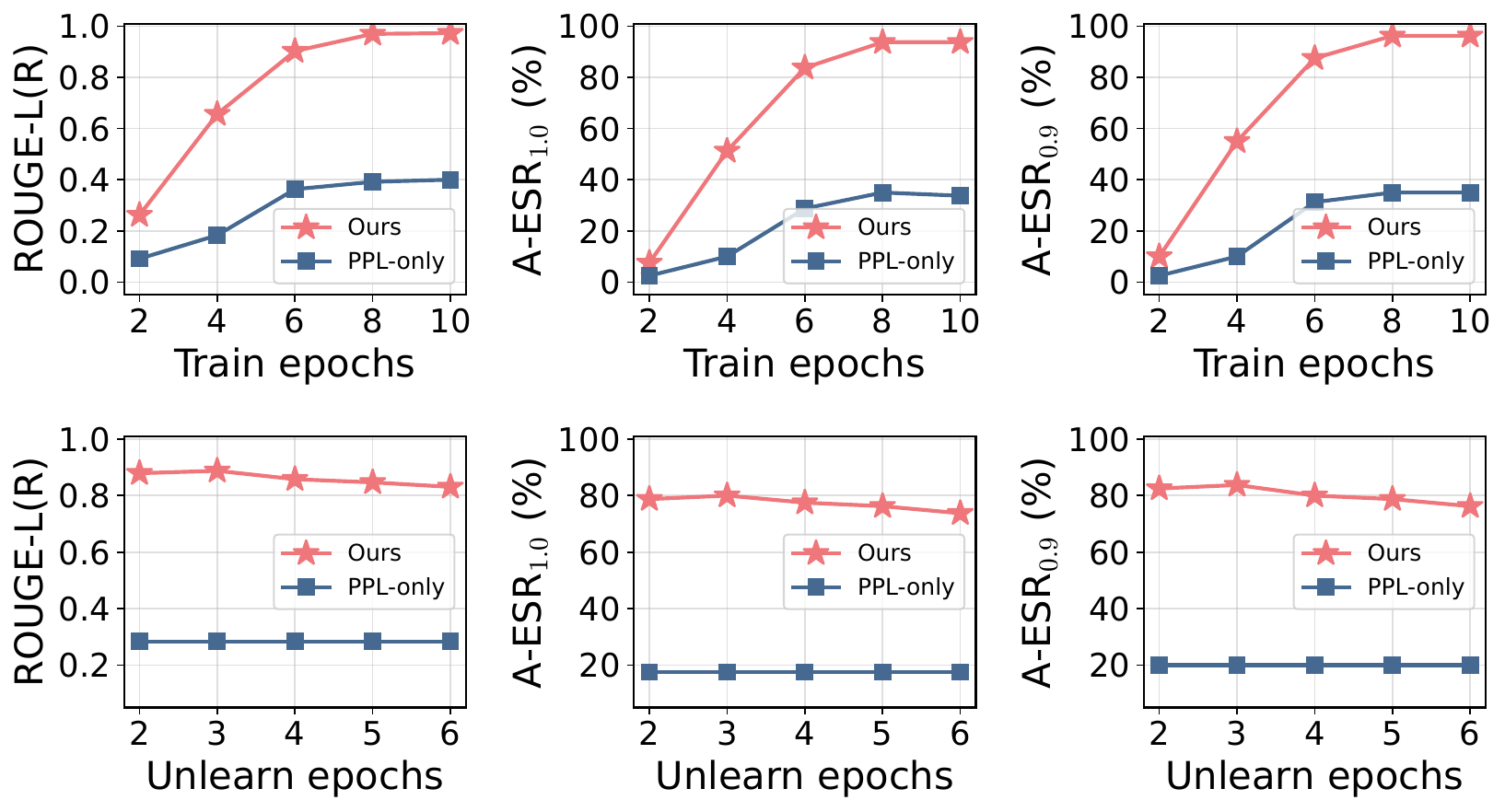}
    \caption{Effect of training and unlearning epochs. Top: varying training epochs (unlearning epochs = 5). Bottom: varying unlearning epochs (training epochs = 5).}
    \label{fig:ablation_epoch}
\vspace{-10pt}
\end{wrapfigure}
Figure~\ref{fig:ablation_epoch} presents the impact of training and unlearning epochs on reconstruction performance on MUSE using TinyLlama under NPO. 
We observe that varying the number of unlearning epochs has a minimal effect on reconstruction performance.
In contrast, increasing the number of training epochs consistently enhances reconstruction quality, suggesting that stronger pre-unlearning memorization makes the target data more recoverable.
Notably, our method TRACE consistently outperforms the baseline across all settings. 

\subsection{TRACE in the Black-Box Setting}
\label{app:blackbox}
Our main experiments consider a white-box setting, where TRACE exploits embedding differences between the pre- and post-unlearning models to construct a compact candidate token set $V_f$. We further investigate whether TRACE can operate under a weaker black-box setting, where model parameters are inaccessible but token-level output probabilities from both model versions are available. In this setting, we therefore set $V_f$ to the entire vocabulary and perform reconstruction. 
We evaluate this setting on MUSE using TinyLlama under NPO with an unlearning batch size of 8, considering pre-unlearning models trained for 4, 5, and 6 epochs.

As shown in Table~\ref{tab:blackbox}, TRACE retains meaningful
reconstruction capability in the black-box setting, although its
performance is lower than in the white-box setting. Without
parameter access, TRACE must instead search over the entire vocabulary,
introducing many competing candidate prefixes and making it more
difficult to retain the ground-truth paths during exploration. This
challenge would become increasingly pronounced for models with larger
vocabularies. In such a large search space, stronger memorization of the target samples exposes them to greater reconstruction risk, as evidenced by the substantial improvement in reconstruction performance with
increasing pre-unlearning training epochs.
\begin{table*}[t]
    \centering
    \caption{TRACE performance in the black-box setting.}
    \label{tab:blackbox}
    \small
    \scalebox{0.88}{
    \begin{tabular}{llcccccc}
        \toprule
        & & \multicolumn{2}{c}{Training epochs = 4} & \multicolumn{2}{c}{Training epochs = 5} & \multicolumn{2}{c}{Training epochs = 6} \\
        \cmidrule(lr){3-4} \cmidrule(lr){5-6} \cmidrule(lr){7-8}
        Dataset & Metric
        & white-box & black-box & white-box
        & black-box & white-box & black-box \\
        \midrule

        \multirow{3}{*}{MUSE}
        & ROUGE-L(R)
        & 0.6566 &  0.3419  & 0.8467
        &0.6043  & 0.9027 & 0.6274 \\
        & A-ESR$_{1.0}$ (\%)
        & 51.25 & 27.50  & 76.25
        & 55.00 & 83.75 & 57.50 \\
        & A-ESR$_{0.9}$ (\%)
        & 55.00 & 27.50 & 78.75
        & 55.00 & 87.50 & 60.00 \\
   
        \bottomrule
    \end{tabular}
    }
\vspace{-10pt}
\end{table*}

\subsection{Robustness to Checkpoint Transformations}
\label{app:checkpoint_transformation}

In practical deployments, model checkpoints may undergo additional
transformations after unlearning. 
We evaluate TRACE under two representative model compression techniques, including low-bit quantization and weight pruning.

\paragraph{Model Quantization.}
\begin{wraptable}{r}{0.35\columnwidth}
    \centering
    \vspace{-10pt}
    \caption{TRACE under 4-bit AWQ quantization.}
    \label{tab:checkpoint_quantization}
    \small
    \renewcommand{\arraystretch}{1.3} 
    \setlength{\aboverulesep}{0pt}
    \setlength{\belowrulesep}{0pt}
    \setlength{\tabcolsep}{4pt}
    \scalebox{0.85}{
    \begin{tabular}{lcc}
        \toprule
        Metric & Original & AWQ 4-bit \\
        \midrule
        ROUGE-L(R)          & 0.9579 & 0.9506 \\
        A-ESR$_{1.0}$ (\%) & 95.00  & 95.00 \\
        A-ESR$_{0.9}$ (\%) & 95.00  & 95.00 \\
        \bottomrule
    \end{tabular}
    }
    \vspace{-8pt}
\end{wraptable}
We apply 4-bit AWQ quantization~\cite{lin2024awq} to the unlearned
LLaMA2-7B model obtained with NPO on MUSE, and evaluate TRACE using the
resulting quantized checkpoint.
As shown in Table~\ref{tab:checkpoint_quantization}, quantization has
only a minor effect on reconstruction performance: ROUGE-L(R) decreases
from 0.9579 to 0.9506, while both A-ESR$_{1.0}$ and A-ESR$_{0.9}$
remain at 95\%. This result shows that the reconstruction signal exploited
by TRACE remains largely preserved after low-bit quantization.

\paragraph{Model Pruning.}
\begin{wraptable}{r}{0.35\columnwidth}
    \centering
    \vspace{-10pt}
    \caption{TRACE under Wanda pruning.}
    \label{tab:checkpoint_pruning}
    \small
\renewcommand{\arraystretch}{1.3} 
    \setlength{\aboverulesep}{0pt}
    \setlength{\belowrulesep}{0pt}
    \setlength{\tabcolsep}{4pt}
    \scalebox{0.85}{
    \begin{tabular}{ccc}
        \toprule
        Pruning Ratio & PPL & ROUGE-L(R) \\
        \midrule
        0\%  & 16.07 & 0.9579 \\
        10\% & 17.50 & 0.9503 \\
        30\% & 13.58 & 0.8489 \\
        50\% & 5.07  & 0.6394 \\
        \bottomrule
    \end{tabular}
    }
    \vspace{-8pt}
\end{wraptable}
We apply Wanda~\cite{sun2024simple} with different pruning ratios to the
unlearned LLaMA2-7B model obtained with NPO on MUSE, and evaluate TRACE
using the resulting pruned checkpoints. 
Table~\ref{tab:checkpoint_pruning} shows that
TRACE remains effective under moderate pruning, with ROUGE-L(R)
decreasing only slightly from 0.9579 to 0.9503 at a 10\% pruning ratio.
More aggressive pruning progressively reduces reconstruction quality,
with ROUGE-L(R) decreasing to 0.8489 and 0.6394 at pruning ratios of
30\% and 50\%, respectively.
Interestingly, aggressive pruning also substantially decreases the
perplexity of the unlearned samples, from 16.07 at 0\% pruning to 5.07 at 50\% pruning. Thus, while aggressive pruning weakens TRACE's
reconstruction performance, 
it simultaneously compromises the effect of unlearning.

%% file: sections/6_defense.tex
\section{Possible Defenses}
We evaluate three practical defenses: \textit{parameter obfuscation}~\cite{abadi2016deep}, \textit{gradient pruning}~\citep{zhu2019deep}, and \textit{post-unlearning fine-tuning}~\cite{hu2024learn}. 

\paragraph{Parameter Obfuscation.}
Inspired by DP-SGD~\cite{abadi2016deep}, we adopt a lightweight parameter obfuscation mechanism during unlearning. At each unlearning step, we first clip gradients to $C=1.2$, and then inject noise proportional to the gradient norm, i.e., $\tilde{g} = g + \sigma \|g\| \frac{z}{\|z\|}, \quad z \sim \mathcal{N}(0, I)$, with hyperparameter $\sigma$ controlling the perturbation strength. We emphasize that this approach is heuristic and does not aim to provide formal differential privacy guarantees.
The injected noise perturbs the embedding parameters across the entire vocabulary, which interferes with our candidate token set $V_f$ extraction. To address this, we apply a simple unsupervised refinement. Specifically, we first select the top-$p\%$ tokens ($p=20$ in our setting) with the largest embedding differences, and then perform 1D clustering (K-means with $k=2$) on their log-scaled differences, taking the cluster with larger mean shifts as $V_f$. 
Table~\ref{tab:obfuscation} reports the trade-off between unlearning effectiveness measured by perplexity (higher is better) and reconstruction performance under different noise levels $\sigma$ on MUSE using TinyLlama under GA. 
As $\sigma$ increases, perplexity on the forget set decreases, indicating weaker unlearning. Meanwhile, the precision of $V_f$ sharply deteriorates when $\sigma \ge 0.15$, so that a large number of noise-induced tokens become indistinguishable from ground-truth tokens. For smaller noise levels ($\sigma \le 0.10$), TRACE remains effective with high precision and strong recall of $V_f$.

\begin{table*}[t]
\centering
\caption{Possible defenses. Rec($\mathcal{V}_f$) and Prec($\mathcal{V}_f$) denote recall and precision over $V_f$ with respect to ground-truth tokens from the unlearned data.}
\label{tab:ab}
\small
\renewcommand{\arraystretch}{1.3}
\setlength{\aboverulesep}{0pt}
\setlength{\belowrulesep}{0pt}
\setlength{\tabcolsep}{4pt}

\begin{subtable}[t]{0.49\textwidth}
\centering
\caption{Parameter Obfuscation.}
\label{tab:obfuscation}
\scalebox{0.7}{
\begin{tabular}{cc|c|cc}
\toprule
\rowcolor{headergray}
$\sigma$ & Perplexity & ROUGE-L(R) & Rec($\mathcal{V}_f$) & Prec($\mathcal{V}_f$) \\
\midrule
0    & 20.17 & 0.8750 & 1      & 1 \\
0.01 & 19.37 & 0.8361 & 0.9551 & 1 \\
0.05 & 18.63 & 0.7226 & 0.9038 & 1 \\
0.10 & 17.81 & 0.5375 & 0.8525 & 1 \\
0.15 & 17.06 & 0.4599 & 0.8846 & 0.0601 \\
0.20 & 16.30 & 0.4062 & 0.8525 & 0.0562 \\
\bottomrule
\end{tabular}
}
\end{subtable}
\hfill
\begin{subtable}[t]{0.49\textwidth}
\centering
\caption{Gradient Pruning.}
\label{tab:pruning}
\scalebox{0.7}{
\begin{tabular}{cc|c|cc}
\toprule
\rowcolor{headergray}
$p$ & Perplexity & ROUGE-L(R) & Rec($\mathcal{V}_f$) & Prec($\mathcal{V}_f$) \\
\midrule
0    & 20.17 & 0.8750 & 1 & 1 \\
0.80 & 25.73 & 0.8750 & 1 & 1 \\
0.85 & 19.43 & 0.8750 & 1 & 1 \\
0.90 & 12.81 & 0.8636 & 1 & 1 \\
0.95 & 5.57  & 0.8636 & 1 & 1 \\
0.99 & 2.29  & 0.8182 & 1 & 1 \\
\bottomrule
\end{tabular}
}
\end{subtable}
\vspace{-5pt}
\end{table*}

\paragraph{Gradient Pruning.}
We evaluate gradient pruning~\citep{zhu2019deep} on MUSE using TinyLlama under GA with pruning ratios $p \in \{0.8, 0.85, 0.9, 0.95, 0.99\}$, where a fraction $p$ of gradient components with the smallest magnitudes are pruned (set to zero) during each update. Pruning has little impact on TRACE, with the extraction of $V_f$ remaining unaffected. As shown in Table~\ref{tab:pruning}, even with an aggressive pruning ratio of $0.99$, TRACE still achieves an average ROUGE-L (R) of 0.8182. 

\paragraph{Fine-tuning.}

\begin{wrapfigure}{r}{0.6\linewidth}
\vspace{-8pt}
  \centering
\includegraphics[width=0.99\linewidth]{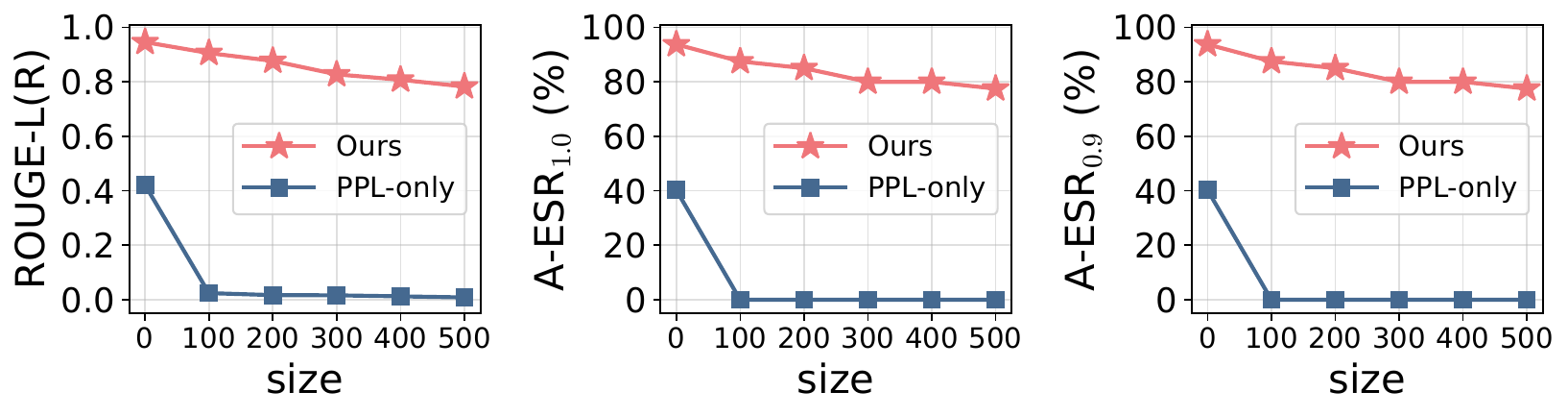}
    \caption{Fine-tuning after unlearning.}
    \label{fig:defense_ft}
\vspace{-8pt}
\end{wrapfigure}

Following the prior work~\cite{hu2024learn}, we consider a setting where the developer fine-tunes the unlearned model before uploading the updated model. Specifically, we fine-tune the unlearned model (obtained via GA) for two epochs on retain sets of varying sizes 0, 100, 200, 300, 400, and 500 on the MUSE dataset using LLaMA2-7B.
As shown in Fig.~\ref{fig:defense_ft}, fine-tuning partially reduces reconstruction performance as the size of the fine-tuning dataset increases. This process enlarges the token candidate set $V_f$ and expands the search space, making reconstruction more challenging.
Nevertheless, TRACE consistently outperforms the PPL-only baseline.

\paragraph{Discussion.}
These results suggest that heuristic defenses provide limited protection against TRACE. Gradient pruning has little effect on reconstruction. Parameter obfuscation can reduce attack effectiveness, but at the cost of weaker unlearning performance; Post-unlearning fine-tuning partially degrades reconstruction by expanding the search space. However, it does not eliminate the perplexity gap induced by unlearning, and TRACE can still be applied, albeit at the cost of increased computational overhead (analyzed in  Appendix~\ref{overhead_compute}), which is often acceptable in offline attack settings. 
We further investigate whether exact retraining can provide stronger
resistance to TRACE in Appendix~\ref{app:retrain}. The results show that exact retraining substantially reduces the attack effectiveness compared with approximate unlearning, although a small fraction of the unlearned samples can still be exactly reconstructed by TRACE.
Overall, these findings underscore the need for unlearning methods and deployment protocols that explicitly account for the attack surface introduced by approximate unlearning.

\section{Conclusion}
In this work, we reveal that approximate unlearning in LLMs leaves recoverable footprints in model snapshots, which are commonly retained for auditing or rollback, facilitating the privacy breaches it is intended to prevent even without requiring any data-specific prior knowledge.
To demonstrate this, we propose TRACE, a framework that reconstructs unlearned text by utilizing parameter differences to effectively constrain the vast search space and employing a contrastive search strategy within a two-stage architecture. Extensive experiments show that TRACE consistently achieves high-fidelity reconstruction across diverse models, datasets, and unlearning methods.
This highlights the need for unlearning methods and deployment protocols that explicitly account for parameter-difference leakage, and call for future work on defenses that mitigate cross-snapshot information exposure.